\documentclass[journal]{IEEEtran}

\usepackage[nosort]{cite}

\ifCLASSINFOpdf
\else
\fi

\usepackage[normalem]{ulem}

\usepackage{amsmath}
\usepackage{algorithm}
\usepackage{algorithmic}
\usepackage{array}
\usepackage{stfloats}
\usepackage{xcolor}
\usepackage{indentfirst}
\usepackage{epsfig,setspace}
\usepackage{float}
\usepackage{bm}
\usepackage{multirow, hhline}
\usepackage{makecell}
\usepackage{amssymb}
\usepackage{pifont}
\usepackage{arydshln}
\usepackage{graphicx}
\usepackage{booktabs}
\newcommand{\cmark}{\ding{51}}
\newcommand{\xmark}{\ding{55}}
\newcommand{\tabincell}[2]{\begin{tabular}{@{}#1@{}}#2\end{tabular}}
\usepackage{subfig}
\usepackage{caption}

\usepackage{changepage}

\newcommand{\rank}{\operatorname{rank}}

\begin{document}
\bstctlcite{IEEEexample:BSTcontrol}

\title{Homogeneous Speaker Features for On-the-Fly Dysarthric and Elderly Speaker Adaptation}

\author{Mengzhe Geng, Xurong Xie, Jiajun Deng, Zengrui Jin, Guinan Li, Tianzi Wang, Shujie Hu, Zhaoqing Li, \\ Helen Meng, Xunying Liu
    \thanks{Mengzhe Geng, Jiajun Deng, Zengrui Jin, Guinan Li, Tianzi Wang, Shujie Hu and Zhaoqing Li are with the Chinese University of Hong Kong, China (email: \{mzgeng,jjdeng,zrjin,gnli,twang,sjhu,zqli\}@se.cuhk.edu.hk).\\
    \indent Xurong Xie is with Institute of Software, Chinese Academy of Sciences, Beijing, China (email: xurong@iscas.ac.cn).\\
    \indent Helen Meng is with the Chinese University of Hong Kong, China (email: hmmeng@se.cuhk.edu.hk).\\
    \indent Xunying Liu is with the Chinese University of Hong Kong, China and the corresponding author (email: xyliu@se.cuhk.edu.hk).\\
    }
}

\markboth{Journal of \LaTeX\ Class Files,~Vol.~14, No.~8, August~2021}%
{Shell \MakeLowercase{\textit{et al.}}: A Sample Article Using IEEEtran.cls for IEEE Journals}


\maketitle

\begin{abstract}

The application of data-intensive automatic speech recognition (ASR) technologies to dysarthric and elderly adult speech is confronted by their mismatch against healthy and non-aged voices, data scarcity and large speaker-level variability. To this end, this paper proposes two novel data-efficient methods to learn homogeneous dysarthric and elderly speaker-level features for rapid, on-the-fly test-time adaptation of DNN/TDNN and Conformer ASR models. These include: 1) speaker-level variance-regularized spectral basis embedding (VR-SBE) features that exploit a special regularization term to enforce homogeneity of speaker features in adaptation; and 2) feature-based learning hidden unit contributions (f-LHUC) transforms that are conditioned on VR-SBE features. Experiments are conducted on four tasks across two languages: the English UASpeech and TORGO dysarthric speech datasets, the English DementiaBank Pitt and Cantonese JCCOCC MoCA elderly speech corpora. The proposed on-the-fly speaker adaptation techniques consistently outperform baseline iVector and xVector adaptation by statistically significant word or character error rate reductions up to 5.32\% absolute (18.57\% relative) and batch-mode LHUC speaker adaptation by 2.24\% absolute (9.20\% relative), while operating with real-time factors speeding up to 33.6 times against xVectors during adaptation. 
The efficacy of the proposed adaptation techniques is demonstrated in a comparison against current ASR technologies including SSL pre-trained systems on UASpeech,
where our best system produces a state-of-the-art WER of 23.33\%.
Analyses show VR-SBE features and f-LHUC transforms are insensitive to speaker-level data quantity in test-time adaptation. T-SNE visualization reveals they have stronger speaker-level homogeneity than baseline iVectors, xVectors and batch-mode LHUC transforms. 

\end{abstract}

\begin{IEEEkeywords}
Dysarthric Speech Recognition, Elderly Speech Recognition, Rapid Adaptation, Speaker Features
\end{IEEEkeywords}

\section{Introduction}
\label{sec:intro}

\IEEEPARstart{I}{n} spite of the remarkable advancement on automatic speech recognition (ASR) techniques for normal speech, accurate recognition of disordered speech, for example, voice recorded from speakers with dysarthria, remains a highly challenging task to date~\cite{christensen2012comparative,sehgal2015model,hu2019cuhk,shahamiri2021speech,baskar2022speaker,hernandez22_interspeech,violeta2022investigating,hu2022exploiting,almadhor2023e2e,kim2023unsupervised,jin2022personalized}. Millions of people worldwide are clinically diagnosed with speech disorders~\cite{lanier2010speech}. In addition,
neurocognitive disorders, such as Alzheimer’s disease (AD), are often found among older adults and manifest themselves in speech impairments~\cite{fraser2016linguistic,alzheimer20192019}. Given that large-scale pathological assessment of speech disorders among all older adults is practically difficult due to the limited availability of professional speech pathologists, the actual number of people affected by speech disorders is much larger. As aging presents enormous challenges to health care worldwide, there is a pressing need to develop suitable ASR-based assistive technologies customized for dysarthric and elderly speakers to improve their life quality and social inclusion~\cite{young2010difficulties,vipperla2010ageing,christensen2013combining,zhou2016speech,vachhani2017deep,kim2018dysarthric,joy2018improving,liu2019exploiting,Shor2019,geng2020investigation,lin2020staged,kodrasi2020spectro,xiong2020source,takashima2020two,liu2021recent,jin2021adversarial,hu2021bayesian,wang2021improved,green2021automatic,ye2021development,pan2021using,geng2021spectro,harvill2021synthesis,geng2022spectro,hu2023exploring,geng2023fly,jin2022adversarial,qi2023parameter,geng2023use,wang2024enhancing}.

Dysarthric and elderly speech bring challenges on three fronts to current ASR technologies predominantly targeting normal speech recorded from healthy, non-aged users. These challenges include: \textbf{a) substantial mismatch against normal speech} due to the underlying neuro-motor control conditions~\cite{jerntorp1992stroke,whitehill2000speech} and aging, for example, imprecise articulation, hoarse voice and increased disfluency~\cite{caligiuri1989influence,peppe2009prosody}; \textbf{b) data scarcity} due to the difficulty in collecting large quantities of such data from dysarthric and elderly speakers with mobility limitations; and \textbf{c) large speaker-level diversity} among dysarthric and elderly talkers, when sources of variability commonly found in normal speech, e.g. accent or gender, are further aggregated with speech pathology severity and aging. 

A key task for all ASR systems is to model the speech variability attributed to speaker-level characteristics. To this end, speaker adaptation techniques provide a key role in customizing ASR systems for users’ needs. For normal speech recognition tasks, three main categories of such techniques have been studied: \textbf{1) auxiliary embedding features} that are speaker-dependent (SD)~\cite{abdel2013fast,saon2013speaker,senior2014improving}; \textbf{2) feature transformations} that produce speaker-independent (SI) features to remove such variability at the front-ends~\cite{gales1998maximum,uebel1999investigation,seide2011feature}; and \textbf{3) model-based adaptation} using specifically designed SD DNN parameters~\cite{gemello2007linear,li2010comparison,swietojanski2016learning,zhang2016dnn}. A detailed review of speaker adaptation techniques targeting normal speech is presented in Sec.~\ref{sec:review}.

Modeling the large speaker-level heterogeneity in dysarthric and elderly speech requires powerful speaker adaptation techniques to be developed. However, only limited prior research in this direction has been conducted. Earlier works focused on deriving speaker adaptation methods for traditional HMM-based ASR systems with Gaussian mixture model (GMM) based hidden state densities, predominantly targeting dysarthric speech only. A combination of HMM state transition interpolation and maximum a posteriori (MAP) adaptation is utilized to account for dysarthric speaker diversity in~\cite{sharma2010state}. The use of maximum likelihood linear regression (MLLR) and MAP adaptation was explored in~\cite{christensen2012comparative,baba2001elderly,mengistu2011adapting,kim2013dysarthric}. Combined use of MLLR with MAP adaptation in speaker adaptative training (SAT) of GMM-HMM models was investigated in~\cite{sehgal2015model}. Feature-space MLLR (f-MLLR) based SAT training~\cite{bhat2016recognition} and regularized speaker adaptation to dysarthric speakers using Kullback-Leibler (KL) divergence~\cite{kim2017regularized} were also developed for GMM-HMM systems. MLLR adaptation of GMM-HMM models to elderly speakers using cross-speaker statistics pooling was studied in~\cite{baba2001elderly}.

More recent dysarthric and elderly speaker adaptation approaches applied to state-of-the-art hybrid and end-to-end (E2E) neural network based ASR systems include the following categories: 1) direct speaker-level parameter fine-tuning of hybrid LF-MMI trained TDNN~\cite{xiong2020source,takashima2020two} and RNN-T~\cite{tobin2022personalized} models; 2) the use of iVector speaker adaptation for dysarthric~\cite{espana2016automatic,yue2022acoustic} and elderly~\cite{ye2021development} speech recognition; 3) model-based speaker adaptation using, for example, learning hidden unit contributions (LHUC) for dysarthric~\cite{geng2020investigation,liu2021recent,jin2021adversarial,geng2022spectro} and elderly~\cite{ye2021development,geng2022spectro} speakers; 4) Bayesian domain~\cite{deng2021bayesian} and speaker adaptation~\cite{liu2021recent} methods that are more robust to dysarthric and elderly speech data scarcity; 5) dysarthric and elderly speaker-level averaged spectro-temporal basis embedding features~\cite{geng2021spectro,geng2022spectro} for hybrid DNN/TDNN and E2E Conformer adaptation; 6) f-MLLR and xVector based dysarthric speaker adaptation of self-supervised learning (SSL) based Wav2vec 2.0 models~\cite{baskar2022speaker}; and 7) dysarthric speaker adapter fusion of pre-trained E2E Transformer models~\cite{qi2023parameter}. 


Suitable dysarthric and elderly speaker adaptation methods that can meet with above-mentioned challenges should satisfy the following requirements: \textbf{1) high data efficiency} to model the very limited speaker-level data; \textbf{2) strong speaker homogeneity} to ensure the distinct speaker-level characteristics are consistently represented in the adapted model; and \textbf{3) low processing latency} to allow adaptation to be performed immediately on-the-fly from the onset of a new user's enrolment in the system to minimize their efforts and fatigue. In this regard, prior research only addressed some of the above. For example, the Bayesian model based adaptation using very limited speaker data~\cite{liu2021recent,ye2021development,deng2021bayesian} only addresses the aforementioned data scarcity issue, but the latency problem remains unvisited. Similarly, the spectro-temporal deep embedding features~\cite{geng2021spectro,geng2022spectro} are averaged over all speaker-level data in an offline manner. This introduces considerable processing latency 
and is thus unsuitable for on-the-fly test-time adaptation.





To this end, one possible solution is to derive suitable rapid, on-the-fly feature-based dysarthric and elderly adaptation techniques. Such methods serve as multi-purpose solutions to handle not only the speaker-level data scarcity and diversity when representing speaker attributes, but also the processing latency incurred by model-based fine-tuning or adaptation to speaker data. In this paper, two novel forms of feature-based on-the-fly rapid speaker adaptation approaches are proposed. The first is based on speaker-level variance-regularized spectral basis embedding (VR-SBE) features. An extra variance regularization term is included when training spectral basis embedding neural networks~\cite{geng2021spectro,geng2022spectro} to ensure speaker homogeneity of the embedding features. This in turn allows them to be applied on-the-fly during test-time adaptation to dysarthric or elderly speakers. The second approach utilizes on-the-fly feature-based LHUC (f-LHUC) transforms conditioned on VR-SBE features. Specially designed regression TDNN~\cite{xie2019fast} predicting dysarthric or elderly speaker-level LHUC transforms is constructed to directly generate and apply such SD parameters on the fly during test-time adaptation. 

Experiments are conducted on four different tasks: 1) the English UASpeech~\cite{kim2008dysarthric} and TORGO~\cite{rudzicz2012torgo} dysarthric speech datasets; 2) the English DementiaBank Pitt~\cite{becker1994natural} and the Cantonese JCCOCC MoCA~\cite{xu2021speaker} elderly speech corpora. Among these, UASpeech is by far the largest available and most extensively used dysarthric speech corpus, while DementiaBank Pitt is the largest publicly available elderly speech database. The performance of the proposed two feature-based on-the-fly speaker adaptation approaches: VR-SBE and VR-SBE feature conditioned f-LHUC, are compared against those of baseline iVector~\cite{saon2013speaker} or xVector~\cite{snyder2018x} based on-the-fly adaptation and LHUC~\cite{swietojanski2016learning} based model adaptation on three fronts:

\textbf{1) ASR performance in WER/CER:} Our proposed on-the-fly speaker adapted hybrid DNN/TDNN and E2E Conformer systems consistently outperform the corresponding on-the-fly iVector/xVector adapted systems, with statistically significant\renewcommand\footnoterule{\vspace{-8pt}}\footnote{Matched pairs sentence-segment word error (MAPSSWE) based statistical significance test~\cite{pallet1990tools} is performed at a significance level $\alpha = 0.05$.} word/character error rate (WER/CER) reductions up to \textbf{5.32\% absolute (18.57\% relative)}. Our on-the-fly adaptation methods also outperform the comparable offline batch-mode model-based LHUC adaptation by statistically significant WER/CER reductions up to \textbf{2.24\% absolute (9.20\% relative)}.

\textbf{2) Analysis on processing latency:} Experiments on the benchmark UASpeech dysarthric and DementiaBank Pitt elderly speech corpora suggest that statistically significant WER/CER reductions are consistently obtained over the corresponding on-the-fly iVector/xVector adapted systems, while the proposed VR-SBE speaker adaptation operates with a real-time factor speeding up ratio up to 33.6 times against xVector. It incurs a minimal processing latency by using an input acoustic feature window \textbf{as short as 10 ms} in duration. 


\textbf{3) Analysis on speaker feature homogeneity:} T-SNE visualization~\cite{van2008visualizing} of baseline iVectors/xVectors, offline model-based LHUC transforms, offline SBE~\cite{geng2022spectro} features, our proposed on-the-fly VR-SBE features and the associated f-LHUC transforms across varying amounts of adaptation data is conducted. This suggests more consistent, data quantity invariant dysarthric and elderly speaker characteristics can be learned via the proposed on-the-fly speaker adaptation approaches.

The main contributions of this paper are summarized below:

1) This paper presents \textbf{novel approaches to learn homogeneous speaker features tailored for rapid, on-the-fly dysarthric and elderly speaker adaptation.} In contrast, prior studies either considered adaptation techniques using all speaker-level data and operating in batch-mode, offline manner~\cite{geng2020investigation,deng2021bayesian,jin2021adversarial,liu2021recent,ye2021development,geng2022spectro,qi2023parameter}, or used existing iVector/xVector features not tailored for dysarthric/elderly speech and produced mixed results on such data~\cite{espana2016automatic,ye2021development,baskar2022speaker,yue2022acoustic}. In particular, model-based adaptation methods not only use all speaker-level data, but also introduce multiple decoding passes and explicit parameter estimation or fine-tuning stages during test-time adaptation~\cite{deng2021bayesian,jin2021adversarial,liu2021recent,ye2021development,qi2023parameter}. 

2) The proposed two on-the-fly feature-based speaker adaptation methods can consistently learn the latent features characterizing dysarthric and elderly speech, such as reduced speech volume and clarity, irrespective of the amount of speaker-level adaptation data available, as demonstrated in the t-SNE visualizations against baseline iVector or xVector adaptation and LHUC model-based adaptation approaches. 


3) Our proposed on-the-fly feature-based adaptation methods produce statistically significant performance improvements over the baseline iVector/xVector adapted
hybrid DNN/TDNN and E2E Conformer systems. Statistically significant WER/CER reductions up to \textbf{5.32\% absolute (18.57\% relative)} on four dysarthric/elderly speech corpora across two languages are obtained, while real-time factor speeding up ratio up to 33.6 times obtained against xVector. In addition, our approaches also produce statistically significant WER/CER reductions up to 2.24\% absolute (9.20\% relative) over LHUC adaptation. These confirm our proposed approaches’ effectiveness and genericity in learning homogeneous speaker features for rapid, on-the-fly dysarthric/elderly speaker adaptation. 

The rest of this paper is structured as follows. Sec.~\ref{sec:review} reviews speaker adaptation targeting normal speech. Sec.~\ref{sec:VR-SBE-all} introduces the proposed on-the-fly VR-SBE features. The associated on-the-fly f-LHUC transforms driven by VR-SBE are presented in Sec.~\ref{sec:f-LHUC-all}. Sec.~\ref{sec:implement} addresses several implementation issues impacting the performance of on-the-fly VR-SBE adaptation and estimation of f-LHUC transforms. The experimental results and analyses are presented in Sec.~\ref{sec:exp}, while Sec.~\ref{sec:conclusion} provides conclusions and future research directions.


\section{Review of Speaker Adaptation}
\label{sec:review}

This section reviews three major categories of speaker adaptation techniques traditionally designed for normal non-aged, healthy speakers, respectively based on: a) auxiliary features, b) feature transformations, and c) model-based adaptation. 

Among these, auxiliary speaker embedding features encode speaker-dependent (SD) characteristics via compact representations. The resulting SD features are used as auxiliary inputs to facilitate speaker adaptation during the training and evaluation of ASR systems. Such SD features can be either estimated independently of the back-end systems, e.g. using universal background models (UBMs) based on Gaussian mixture models (GMMs) to learn iVectors~\cite{saon2013speaker,senior2014improving}, or jointly learned with the back-end systems, e.g. alternate updating speaker representations and the remaining parameters when learning speaker codes~\cite{abdel2013fast}. Auxiliary speaker features can be flexibly incorporated into both GMM-HMM or hybrid systems and more recent end-to-end E2E systems~\cite{tuske2021limit,zeineldeen2022improving,baskar2022speaker}.

Speaker adaptative feature transformations are applied to acoustic front-ends to produce speaker-invariant canonical input features, for example, feature-space maximum likelihood linear regression (f-MLLR)~\cite{gales1998maximum} estimated at speaker-level from GMM-HMM systems. Speaker-level physiological differences motivated adaptation based on vocal tract length normalization (VTLN)~\cite{eide1996parametric,lee1996speaker} can be further adopted by applying piecewise linear frequency warping factors~\cite{lee1996speaker}. 



Model-based speaker adaptation techniques apply compact forms of specially designed SD parameters in different network layers, for example, linear input networks (LIN)~\cite{neto1995speaker,li2010comparison}, linear hidden networks (LHN)~\cite{gemello2007linear}, learning hidden unit contributions (LHUC)~\cite{swietojanski2016learning,xie2021bayesian}, linear output networks (LON)~\cite{gemello2007linear}, parameterized activation functions (PAct)~\cite{zhang2016dnn}, factorized linear transformation\cite{zhao2016low}, and SD neural beamforming, encoder, attention or decoder modules in E2E multi-channel systems~\cite{ochiai2018speaker}. Speaker adaptive training (SAT)~\cite{anastasakos1996compact,ochiai2014speaker} can be further applied during training to enable a joint optimization of both SD and SI parameters and to produce a canonical model that can be better adapted to unseen speakers during test-time adaptation. To alleviate the risk of over-fitting to limited speaker data during model adaptation, a series of regularized speaker adaptation strategies have been developed. These include the use of L2~\cite{liao2013speaker,kitza2018comparison} or Kullback–Leibler (KL) divergence regularization~\cite{yu2013kl,li2018speaker,meng2019speaker,huang2020acoustic}, and maximum a posterior (MAP)~\cite{huang2015maximum,huang2016bayesian} or Bayesian inspired adaptation~\cite{xie2021bayesian}. Data augmentation on the target speaker via text-to-speech (TTS) synthesis has also been explored in ~\cite{huang2020rapid,sim2019personalization}. Alternative objective functions have been investigated for speaker adaptation, for example, adversarial learning ~\cite{meng2018speaker,tsuchiya2018speaker,meng2019speaker} and multitask learning~\cite{price2014speaker,swietojanski2015structured,huang2015rapid,li2018speaker,meng2019speaker}.

\section{On-the-Fly Speaker Adaptation Via Variance Regularized Spectral Basis Embedding Features}
\label{sec:VR-SBE-all}

\begin{figure*}[htbp]
  \centering
  \includegraphics[scale=0.72]{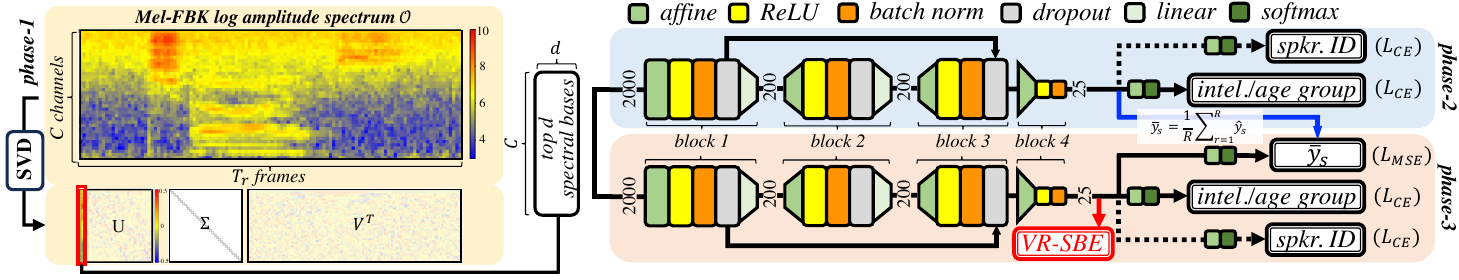}
  \caption{Example extraction of variance-regularized spectral basis embedding (VR-SBE) features (bottom right) for on-the-fly speaker adaptation over 3 phases: \textbf{1)} utterance-level SVD spectrum decomposition (left, in light yellow) with $C = 40$ mel-filterbank channels and retaining top $d = 2$ principal spectral bases; \textbf{2)} multitask learning to derive utterance-level spectral basis embeddings using speaker (``spkr.'') IDs and speech intelligibility (``intel.'') or age groups; and \textbf{3)} extraction of VR-SBE features with averaged speaker embeddings from phase-2 for computing the additional MSE cost. }
\vspace{-\baselineskip} 
  \label{fig:VR-SBE-embedding}
\end{figure*}

On-the-fly feature-based speaker adaptation techniques provide practical solutions to handle both speaker-level data scarcity and processing latency incurred by model-based fine-tuning or adaptation to user data. A key task in designing such techniques is to ensure the homogeneity of speaker-level features and consistent representation of speaker characteristics. To this end, the proposed variance-regularized spectral basis embedding (VR-SBE) features are extracted in three phases (Fig.~\ref{fig:VR-SBE-embedding}), including: \textbf{a)} conducting singular value decomposition (SVD) on utterance-level dysarthric/elderly speech spectra to produce initial time-invariant spectral bases (Fig.~\ref{fig:VR-SBE-embedding} left, in yellow); \textbf{b)} feeding the spectral bases through a first embedding module trained using speaker ID and speech intelligibility/age information, which extracts latent embeddings more consistent and relevant to dysarthric/elderly speaker attributes (Fig.~\ref{fig:VR-SBE-embedding} top right, in light blue); and \textbf{c)} feeding the spectral bases through a second embedding module, with an additional output variance regularization cost using the averaged speaker embeddings obtained in stage b) as targets (Fig.~\ref{fig:VR-SBE-embedding} bottom right, in pink). Such a process ensures the maximum speaker homogeneity of the final VR-SBE features for on-the-fly adaptation.

\vspace{-0.5em}
\subsection{Speech Spectrum Subspace Decomposition}
\label{sec:SVD}


SVD-based spectrum decomposition~\cite{geng2021spectro,geng2022spectro} provides an intuitive approach to decouple the latent time-invariant spectral and time-variant temporal features in speech signals. In phase-1 of VR-SBE feature extraction (Fig.~\ref{fig:VR-SBE-embedding} left, in yellow), SVD is applied to the mel-filterbank log amplitude spectrum $\bm{{\mathcal{O}}}^{r}_{C \times T_r}$ of utterance $r$ with $T_r$ frames and $C$ filterbank channels:

\vspace{-1em}
\setlength{\belowdisplayskip}{0pt} \setlength{\belowdisplayshortskip}{0pt}
\setlength{\abovedisplayskip}{0pt} \setlength{\abovedisplayshortskip}{0pt}
\begin{equation}
    \label{eq:VR-SBE_svd}
    \bm{\mathcal{O}}^{r}_{C \times T_r} = \bm{U}_{r}\bm{\Sigma}_{r}\bm{V}_{r}^{\mathrm{T}}
\end{equation}
\vspace{-1em}


\noindent where the set of column vectors of the $C \times C$ dimensional $\bm{U}_r$ matrix (the left-singular vectors) and the set of row vectors of the $T_{r} \times T_{r}$ dimensional $\bm{V}_{r}^{\mathrm{T}}$ matrix (the right-singular vectors) are respectively the bases of the spectral and temporal subspaces.  $\Sigma_{r}$ is a $C \times T_r$ diagonal matrix containing the singular values in descending order. Its rank equals the number of non-zero singular values, i.e., $\rank(\bm{\mathcal{O}}^{r}_{C \times T_r}) \leq \min \{C, T_r\}$. 

The top few, for example, $d=2$, spectral bases are found to contain the most distinct time-invariant dysarthric or elderly speech features for speaker adaptation, such as overall reduced speech volume and clarity, when compared with retaining all spectral bases or further using the temporal bases~\cite{geng2021spectro,geng2022spectro}. 

To tailor for the low latency requirement of on-the-fly adaptation tasks, instead of performing SVD on the complete spectrum of each utterance, such decomposition can also be performed on part of each utterance’s spectrum in a streaming mode via a sliding analysis widow, e.g. as short as 10 ms. Detailed ablation studies on the number of principal spectral bases and the size of the spectrum decomposition context window are shown in Sec.~\ref{sec:dim} and~\ref{sec:exp} (Tables~\ref{tab:VR-SBE-sliding-window-UASpeech} and~\ref{tab:VR-SBE-sliding-window-DBANK}). 

\vspace{-0.5em}
\subsection{Spectral Basis Deep Embedding}
\label{sec:SBE}

The above spectral bases are obtained in an unsupervised manner during SVD. To further extract latent information more consistent and relevant to speech impairment severity or age information, supervised feature learning is performed in phase-2 (Fig. 1 top right, in light blue) via developing a 4-block speech intelligibility or age embedding module with the selected $top\ d$ spectral bases as inputs. The first three blocks each contain 2000 hidden nodes, while the fourth has a 25-dim bottleneck for embedding extraction. Each block comprises the following internal structures in sequence: affine transformation (in green), rectified linear unit (ReLU) activation (in yellow), and batch normalization (in orange). Additionally, we apply linear bottleneck projection (in light green), dropout operation (in grey) and softmax activation (in dark green) respectively to the intermediate two blocks' inputs, the first three blocks' outputs and the output block. The first block is connected to the third via a skipping connection. Following~\cite{geng2022spectro}, the training targets comprise speech intelligibility groups + speaker IDs for dysarthric speech and binary aged vs. non-aged annotations for elderly speech. We then extract 25-dim spectral basis embedding (SBE) features from the trained bottleneck block. 


\vspace{-0.5em}
\subsection{Variance-Regularized Deep Embedding}
\label{sec:VR-SBE}

Feature-based adaptation techniques require speaker homogeneity to be consistently encoded in the spectral basis embedding features obtained above in Sec.~\ref{sec:SBE}. As such features are computed over individual utterances or portions of them, additional smoothing is required to ensure their homogeneity, e.g. an overall reduction of speech volume among dysarthric or elderly speakers. Prior research applies an offline speaker-level averaging~\cite{geng2022spectro} of all utterance-level produced features, which conflicts with the low processing latency and streaming objective of on-the-fly adaptation. 


To this end, an alternative form of speaker-level spectral basis embedding features smoothing based on variance regularization is proposed. This requires a second embedding module to be constructed in phase-3 (Fig.~\ref{fig:VR-SBE-embedding} bottom right, in pink). During its training, an additional regression task is adopted using the speaker-level averaged spectral basis embeddings obtained from phase-2 as targets (Fig.~\ref{fig:VR-SBE-embedding} blue bold line), which minimizes the output features’ variance and thus further maximizes speaker homogeneity in the final embedding features for on-the-fly adaptation. The overall multitask learning cost interpolates: 1) the cross-entropy (CE) loss over speech intelligibility/age groups, optionally plus that over speaker IDs, and 2) the mean squared error (MSE) between the bottleneck hidden features and the averaged speaker representations obtained from  phase-2, given as\renewcommand\footnoterule{\vspace{-8pt}}\footnote{The weights are empirically set as $\beta_1=\beta_2=\beta_3=\frac{1}{3}$ for dysarthric speech datasets, while for elderly speech corpora as $\beta_1=\beta_2=\frac{1}{2},\beta_3=0$.}:

\vspace{-1em}
\setlength{\belowdisplayskip}{4pt} \setlength{\belowdisplayshortskip}{4pt}
\setlength{\abovedisplayskip}{0pt} \setlength{\abovedisplayshortskip}{0pt}
\begin{equation}
    \label{eq:VR-SBE_MTL}
    \mathcal{L}_{MTL}={\beta_1}{\cdot}\mathcal{L}_{MSE}+{\beta_2}{\cdot}\mathcal{L}_{CE_{group}}+{\beta_3}{\cdot}\mathcal{L}_{CE_{ID}}
\end{equation}


Here $\mathcal{L}_{MSE}=\frac{1}{R}\sum_{r=1}^{R}(y_{s}^{r}-\overline{y}_{s})^2$, where $y_{s}^{r}$ and $\overline{y}_{s}$ respectively denote the $r^{th}$ bottleneck hidden feature and averaged representation of speaker $s$ from phase-2.



\begin{figure*}[ht]
  \centering
    \subfloat[Adaptation in \textbf{Hybrid TDNN Systems}]{\includegraphics[width=0.48\textwidth]{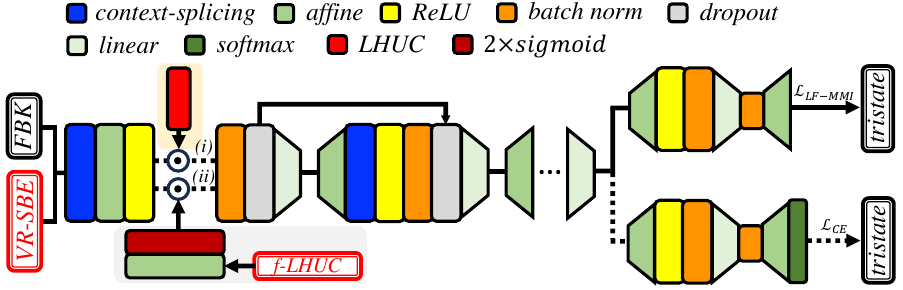}\label{fig:adapt_tdnn}}  
    \hfill
    \subfloat[Adaptation in \textbf{E2E Conformer Systems}]{\includegraphics[width=0.48\textwidth]{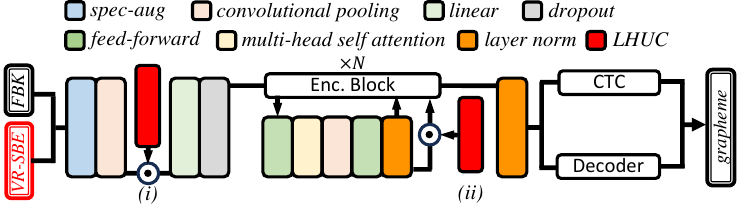}\label{fig:adapt_conformer}}
  \caption{Incorporatiing our on-the-fly VR-SBE adaptation at front-ends of \textbf{(a)} hybrid TDNN and \textbf{(b)} E2E Conformer. \textbf{In (a)},  path (i) leads to additional model-based LHUC adaptation (top) while path (ii) incorporates our on-the-fly f-LHUC adaptation (bottom). \textbf{In (b)}, path (i) and (ii) respectively apply LHUC adaptation after convolution pooling and a particular encoder block.}
   \vspace{-\baselineskip} 
  \label{fig:VR-SBE-adapt}
\end{figure*}

The 25-dim variance-regularized spectral basis embedding (VR-SBE) features are then extracted from the bottleneck of the second embedding module (Fig.~\ref{fig:VR-SBE-embedding} bottom right, red bold line) and concatenated with acoustic features before being fed into hybrid DNN/TDNN (Fig.~\ref{fig:VR-SBE-adapt}(a)) or E2E Conformer (Fig.~\ref{fig:VR-SBE-adapt}(b)) ASR systems to facilitate on-the-fly speaker adaptation during both model training and test-time adaptation.

\section{On-The-Fly Speaker Adaptation via VR-SBE feature driven F-LHUC Transforms}
\label{sec:f-LHUC-all}


Model-based speaker adaptation techniques offer powerful, fine-grained parameterization of speaker-level attributes when personalizing ASR systems. However, their application to dysarthric and elderly speech is often hindered by both the scarcity of speaker-level data and processing latency during user fine-tuning.  In order to leverage the strengths of both model-based and feature-based adaptation techniques, a novel form of on-the-fly model-based adaptation approach using VR-SBE feature-based LHUC (f-LHUC) transformations is proposed in this paper. The following Sec.~\ref{sec:batch-LHUC} reviews standard model-based LHUC speaker adaptation before VR-SBE feature-based LHUC adaptation is introduced in Sec.~\ref{sec:f-LHUC}.


\vspace{-0.5em}
\subsection{Batch-Mode Model-Based LHUC Adaptation}
\label{sec:batch-LHUC}



The standard approach of applying LHUC speaker adaptation involves training speaker-dependent (SD) linear scaling transforms to adjust activation amplitudes on DNN nodes~\cite{swietojanski2014learning,swietojanski2016learning}. Let $\bm{A}^{s}$ denote such transform in a specific layer for speaker $s$. To reduce the risk of overfitting to limited speaker data, $\bm{A}^{s}$ can be further constrained as a diagonal matrix, equivalent to applying vector $\bm{v}^{s}$ to modify amplitudes of activations. The adapted outputs can be expressed as:


\vspace{-1em}
\setlength{\belowdisplayskip}{0pt} \setlength{\belowdisplayshortskip}{0pt}
\setlength{\abovedisplayskip}{0pt} \setlength{\abovedisplayshortskip}{0pt}
\begin{equation}
    \label{eq:standard_lhuc}
    \bm{h}^{s}=\xi(\bm{v}^{s}) \odot \bm{h}
\end{equation}



\noindent where $\bm{h}$ and $\odot$ denote the activated hidden vector and the Hadamard product operation. $\xi(\cdot)$ denotes the activation function, e.g. the element-wise $2 \times sigmoid(\cdot)$ function~\cite{swietojanski2016learning} ranging from 0 to 2. In this case, the SD LHUC parameters $\bm{v}^{s}$ are typically initialized as $\bm{0}$. Such parameters can be estimated by finding the minimum cross entropy (CE) estimator $\bm{\tilde{v}}^{s}_{CE}$:


\vspace{-1em}
\setlength{\belowdisplayskip}{0pt} \setlength{\belowdisplayshortskip}{0pt}
\setlength{\abovedisplayskip}{0pt} \setlength{\abovedisplayshortskip}{0pt}
\begin{equation}
    \label{eq:LHUC_CE}
    \tilde{{\bm{v}}}^{s}_{CE} = \arg\min_{{\bm{v}}^{s}} \{ - \log P(\mathcal{H}^{s}|{\bm{\mathcal{D}}}^{s},{\bm{v}}^{s}) \}
    \setlength{\belowdisplayskip}{-1em}
\end{equation}
\vspace{-1em}

\noindent where $\bm{\mathcal{D}}^{s}$ and $\mathcal{H}^{s}$ denote the adaptation data recorded from speaker $s$ and the corresponding supervision label, e.g. HMM states in hybrid systems or output tokens in E2E systems. Without loss of generality, we omit the SI model parameters.


During unsupervised test-time adaptation, the label $\mathcal{H}^{s}$ is commonly produced by initially decoding the adaptation data $\bm{\mathcal{D}}^{s}$. Such batch-mode adaptation process can be iteratively performed to refine both the quality of hypothesis supervision and resulting re-estimated SD parameters. The adapted model with the final estimated SD parameters undergoes another decoding pass to infer hypotheses $\tilde{\mathcal{H}}^{s}$ for test data $\tilde{\bm{\mathcal{D}}}^{\raisebox{-2pt}{$\scriptstyle s$}}$.



When using LHUC-based speaker adaptive training (LHUC-SAT)~\cite{anastasakos1996compact}, the SD parameters tied to training speakers are joint-optimized with the shared SI parameters. The resulting SI parameters provide a canonical model that more effectively learns speaker-invariant speech properties and can be better adapted to unseen speakers during test-time adaptation.


\subsection{Feature-Based LHUC Transforms}
\label{sec:f-LHUC}


All batch-mode speaker adaptation techniques, including those based on LHUC of Sec.~\ref{sec:batch-LHUC}, operate with multiple decoding and SD parameter estimation passes. This normally requires all speaker-level data to be used to ensure the robustness and fine-granularity of SD parameters. In order to improve data efficiency, reduce processing latency and minimize the impact of supervision errors, an alternative approach is to directly learn a homogeneous mapping between speaker data and associated SD parameters, bypassing the multi-pass decoding followed by unsupervised adaptation procedure entirely. This allows speaker-level feature-based LHUC (f-LHUC) transforms to be directly predicted on the fly during evaluation using the speech data arriving in a streaming mode. To ensure the homogeneity of such regression mapping, the VR-SBE features proposed in Sec.~\ref{sec:VR-SBE-all} are used as the regression inputs together with mel-scale filterbank (FBK).

\begin{figure}[htbp]
  \centering
  \includegraphics[scale=0.46]{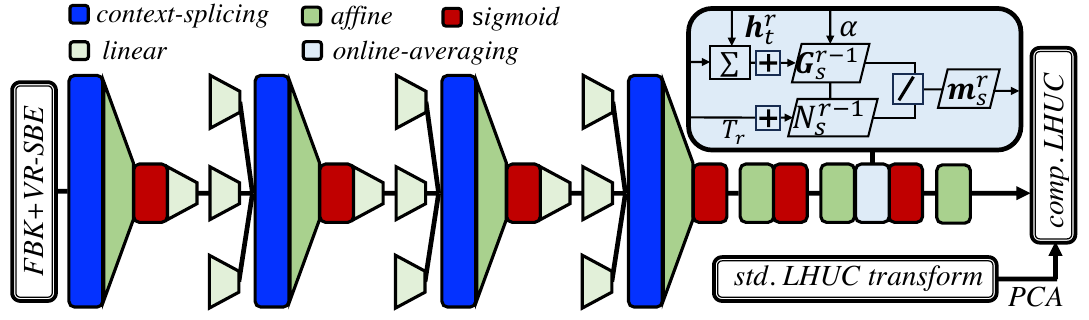}
  \caption{Example f-LHUC regression network using FBK + VR-SBE input features with an online cross-utterance hidden context averaging layer (top right in light blue) to predict homogeneous speaker-level LHUC transforms on the fly. }
  \vspace{-0.5\baselineskip} 
  \label{fig:f-LHUC_regression_network}
\end{figure}  

As shown in Fig.~\ref{fig:f-LHUC_regression_network}, our TDNN LHUC regression network \cite{xie2019fast} contains four 500-dim context-splicing layers (in blue) and two 500-dim feed-forward layers comprising affine transformation (in green) and sigmoid activation (in red). Three 300-dim linear bottleneck projections (in light green) are inserted between the context-splicing layers to reduce the number of parameters, while a specifically designed online averaging layer (in light blue) is integrated before the sigmoid activation of the second feed-forward layer (Fig.~\ref{fig:f-LHUC_regression_network} upper right). Such a layer stores the preceding speech utterances in memory via an accumulated history vector and a frame counter for each speaker. The output of this online averaging layer is given by:

\setlength{\belowdisplayskip}{4pt} \setlength{\belowdisplayshortskip}{4pt}
\setlength{\abovedisplayskip}{-2pt} \setlength{\abovedisplayshortskip}{-2pt}
\begin{equation}
    \label{eq:f-LHUC_calculation}
    \bm{m}_{s}^{r}=\frac{\sum_{t=1}^{T_{r}}\bm{h}_{t}^{r}+{\alpha}{\times}\bm{G}_{s}^{r-1}}{T_{r}+{\alpha}{\times}N_{s}^{r-1}}
\end{equation}

\noindent where $\bm{G}_s^{r-1}$ and $N_s^{r-1}$ refer to the accumulated history vector and the frame counter until the $(r-1)^{th}$ utt. of speaker $s$. $T_{r}$ denotes the frame counts in the $r^{th}$ utt., while $\bm{h}_t^{r}$ denotes its $t^{th}$ hidden vector. $\alpha$ is the history interpolation factor ranging from $[0,1]$. The numerator of Eqn.~(\ref{eq:f-LHUC_calculation}) represents the weighted sum of hidden vectors until the $r^{th}$ utt. of speaker $s$, while the denominator denotes the weighted frame counts. The output $\bm{m}_s^{r}$ represents the averaged representation for speaker $s$ until the $r^{th}$ utt. For all speakers, we set $\bm{G}_s^{0}=\bm{0}$ and $N_s^{0}=0$.


\begin{figure}[htbp]
  \centering
  \includegraphics[scale=0.58]{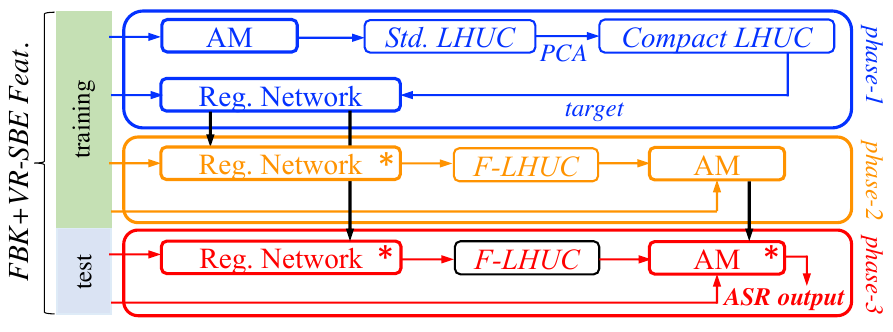}
  \caption{Overall procedure of building VR-SBE feature driven f-LHUC SAT DNN/TDNN systems over three phases: \textbf{1)} f-LHUC regression network training (in blue); \textbf{2)} acoustic model (AM) fine-tuning with f-LHUC transforms (in orange); and \textbf{3)} on-the-fly speaker adaptation with f-LHUC transforms (in red). Here ``AM’’ stands for DNN/TDNN acoustic model, while ``Reg. Network” denotes the f-LHUC regression network. $``\ast"$ indicates the model or network is frozen during inference.}
  \vspace{-\baselineskip} 
  \label{fig:f-LHUC_procedure}
\end{figure}







Fig. 4 demonstrates the three-phase procedure to build our VR-SBE feature driven f-LHUC SAT DNN/TDNN systems: 

\noindent \textbf{Phase-1 f-LHUC regression network training} involves DNN/TDNN acoustic models (Fig.~\ref{fig:adapt_tdnn}) to generate standard (std.) LHUC transforms for each training speaker, before being principal component analysis (PCA) projected to a more compact subset of most distinct bases to serve as the f-LHUC regression (reg.) network (Fig.~\ref{fig:f-LHUC_regression_network}) training targets.

\noindent \textbf{Phase-2 acoustic model fine-tuning} using training speaker f-LHUC transforms inferred by the well-trained f-LHUC regression network of Phase-1 that is now frozen. 

\noindent \textbf{Phase-3 on-the-fly test-time speaker adaptation} of the well-trained DNN/TDNN (Phase-2) using test speaker f-LHUC transforms inferred by the f-LHUC regression network (Phase-1) to produce the speaker-dependent (SD) decoding outputs.

\section{Implementation Details}
\label{sec:implement}


This section discusses several implementation details that affect the performance of our VR-SBE features and f-LHUC transforms, including: \textbf{1)} the number of principal spectral bases as inputs to VR-SBE embedding network; \textbf{2-4)} the inputs to f-LHUC regression network,  its PCA compressed regression targets' dimensionality and its history interpolation factor; and \textbf{5)} baseline speaker adaptation of E2E Conformer. Ablation studies are conducted on the benchmark UASpeech~\cite{kim2008dysarthric} dysarthric and DementiaBank Pitt~\cite{becker1994natural} elderly speech datasets\renewcommand\footnoterule{\vspace{-8pt}}\footnote{A 130.1h training set and a 9h evaluation set are used for UASpeech~\cite{geng2020investigation}, while a 58.9h training set, a 2.5h development set and a 0.6h evaluation set are used for DementiaBank Pitt~\cite{ye2021development}.}. 40-dim mel-filterbank (FBK) log amplitude spectra serve as inputs to SVD. The history interpolation factor $\alpha$ for f-LHUC regression\renewcommand\footnoterule{\vspace{-8pt}}\footnote{The context-splicing indices for f-LHUC regression are set as $\{-2,-1,0,1,2\}$,$\{-2,0,2\}$,$\{-3,0,3\}$ and $\{-4,0,4\}$, following~\cite{xie2019fast}} (Eqn.~(\ref{eq:f-LHUC_calculation}), Sec.~\ref{sec:f-LHUC}) is empirically set to 0.9 unless otherwise stated.

\vspace{-0.5em}
\subsection{Number of Spectral Bases for VR-SBE}
\label{sec:dim}

\begin{table}[htbp]
    \vspace{-10pt}
    \caption{Ablation study of the number of principal spectral bases fed into the VR-SBE embedding network (Fig.~\ref{fig:VR-SBE-embedding}) on \textbf{UASpeech}. ``$d$'' refers to the number of principal spectral vectors. ``VL'', ``L'', ``M'' and ``H'' denote speech intelligibility ``very low'', ``low'', ``mid'' and ``high''. }
    \label{tab:ablation-UASpeech-dim}
    \centering
    \renewcommand\arraystretch{1.0}
    \renewcommand\tabcolsep{2.0pt}
    \scalebox{0.76}{\begin{tabular}{c|c|c|c|c|cccc|c}
    \hline\hline  
        \multirow{2}{*}{Sys.} & 
        \multirow{2}{*}{\tabincell{c}{Model\\(\#Para.)}} & 
        \multirow{2}{*}{\#Hrs} &
        \multicolumn{2}{c|}{Embed. Input} & 
        \multicolumn{5}{c}{WER\%} \\ 
    \cline{4-10}
        & & & $d$ & Dim. & VL & L & M & H & All \\
    \hline\hline
        1 & \multirow{8}{*}{\tabincell{c}{Hybrid\\DNN\\(6M)}} & \multirow{8}{*}{130.1} & 1 & 40 & 62.83 & 29.86 & 20.50 & 9.18 & 28.08 \\
        2 & & & 2 & 80 & \textbf{62.54} & 30.22 & \textbf{18.54} & \textbf{8.59} & \textbf{27.54} \\
        3 & & & 3 & 120 & 63.42 & 29.94 & 19.92 & 8.80 & 27.98 \\
        4 & & & 4 & 160 & 63.79 & 29.13 & 20.96 & 9.25 & 28.20 \\
        5 & & & 5 & 200 & 63.22 & 29.92 & 20.43 & 9.08 &  28.12 \\
        6 & & & 10 & 400 & 62.86 & 29.46 & 19.62 & 9.04 & 27.76 \\
        7 & & & 20 & 800 & 63.52 & 30.06 & 19.76 & 9.44 & 28.22 \\
        8 & & & 40 & 1600 & 64.29 & 31.02 & 22.47 & 9.54 & 29.18 \\
    \hline\hline     
    \end{tabular}}
    \vspace{-5pt}
\end{table}

When performing VR-SBE adaptation of Sec.~\ref{sec:VR-SBE}, Table~\ref{tab:ablation-UASpeech-dim} suggests selecting top-2 SVD derived spectral bases (Fig.~\ref{fig:VR-SBE-embedding}, left) for UASpeech produces the best performance across most speech intelligibility groups, including the most challenging ``VL''. In contrast, Table~\ref{tab:ablation-dbank-dim} shows that the optimal number of principal spectral bases for DementiaBank Pitt is 3. Such settings are respectively adopted for dysarthric and elderly speech datasets in the following VR-SBE adaptation experiments.

\begin{table}[htbp]
    \vspace{-0pt}
    \caption{Ablation study of the number of principal spectral bases fed into the VR-SBE embedding network (Fig.~\ref{fig:VR-SBE-embedding}) on \textbf{DementiaBank Pitt}. ``Dev'' and ``Eval'' refer to the development and evaluation sets. ``INV'' and ``PAR'' denote non-aged clinical investigators and elderly participants.}
    \label{tab:ablation-dbank-dim}
    \centering
    \renewcommand\arraystretch{1.0}
    \renewcommand\tabcolsep{2.0pt}
    \scalebox{0.69}{\begin{tabular}{c|c|c|c|c|cc|cc|c}
    \hline\hline  
        \multirow{3}{*}{Sys.} & 
        \multirow{3}{*}{\tabincell{c}{Model\\(\#Para.)}} & 
        \multirow{3}{*}{\#Hrs} &
        \multicolumn{2}{c|}{Embed. Input} & 
        \multicolumn{5}{c}{WER\%} \\  
    \cline{4-10}
        & & & \multirow{2}{*}{$d$} & \multirow{2}{*}{Dim} & \multicolumn{2}{c|}{Dev} & \multicolumn{2}{c|}{Eval} & \multirow{2}{*}{All (PAR)} \\
    \cline{6-9}
        & & & & & INV & PAR & INV & PAR & \\
    \hline\hline
        1 & \multirow{8}{*}{\tabincell{c}{Hybrid\\TDNN\\(18M)}} & \multirow{8}{*}{58.9} & 1 & 40 & 18.84 & 44.77 & 18.09 & 34.10 & 31.63 (41.61) \\
        2 & & & 2 & 80 & 19.15 & 45.16 & 17.20 & 33.53 & 31.79 (41.72) \\
        3 & & & 3 & 120 & 18.72 & \textbf{44.67} & 18.65 & \textbf{34.03} & \textbf{31.55 (41.52)} \\
        4 & & & 4 & 160 & 18.91 & 44.79 & 18.31 & 34.68 & 31.78 (41.80) \\
        5 & & & 5 & 200 & 19.47 & 45.45 & 18.65 & 34.87 & 32.31 (42.32) \\
        6 & & & 10 & 400 & 19.75 & 45.79 & 20.64 & 34.96 & 32.63 (42.58) \\
        7 & & & 20 & 800 & 19.94 & 45.08 & 18.42 & 34.22 & 32.24 (41.86) \\
        8 & & & 40 & 1600 & 19.50 & 45.17 & 18.42 & 33.53 & 31.98 (41.72) \\
  \hline\hline   
  \end{tabular}}
  \vspace{-5pt}
\end{table}

\vspace{-0.5em}
\subsection{Inputs to F-LHUC Regression Network}
\label{sec:f-LHUC_regression_input}

When performing f-LHUC online speaker adaptation of Sec.~\ref{sec:f-LHUC}, the results in Tables~\ref{tab:ablation-UASpeech-hybrid} and~\ref{tab:ablation-DBANK-hybrid} suggest that on both the UASpeech and DementiaBank Pitt, the combined use of both FBK and VR-SBE features as the inputs to the f-LHUC regression network (Fig.~\ref{fig:f-LHUC_regression_network}, left) generally outperforms using either feature alone, especially when the most powerful configuration combining VR-SBE and f-LHUC is used (Sys.8 vs. Sys.6-7 in Tables~\ref{tab:ablation-UASpeech-hybrid} and~\ref{tab:ablation-DBANK-hybrid}). Hence, both FBK and VR-SBE features are used as the inputs to the f-LHUC regression network for all the following f-LHUC adaptation experiments.

\begin{table}[htbp]
    \vspace{0pt}
    \caption{Ablation study on the input features to the f-LHUC regression network on \textbf{UASpeech}. ``VR-SBE'' denotes variance-regularized spectral basis embedding.}
    \label{tab:ablation-UASpeech-hybrid}
    \centering
    \renewcommand\arraystretch{1.0}
    \renewcommand\tabcolsep{2.0pt}
    \scalebox{0.72}{\begin{tabular}{c|c|c|c|cccc|c}
    \hline\hline  
        \multirow{2}{*}{Sys.} & 
        \multirow{2}{*}{\tabincell{c}{Model\\(\#Para.)}} & 
        \multirow{2}{*}{\tabincell{c}{Adapt.\\Feat.}} & 
        \multirow{2}{*}{f-LHUC} &
        \multicolumn{5}{c}{WER\%} \\  
    \cline{5-9}
        & & & & VL & L & M & H & All \\
    \hline\hline
        1 & \multirow{8}{*}{\tabincell{c}{Hybrid\\DNN\\(6M)}} & \xmark & \multirow{2}{*}{\xmark} & 66.45 & 28.95 & 20.37 & 9.62 & 28.73 \\
        2 & & VR-SBE & ~ & 62.54 & 30.22 & 18.54 & 8.59 & 27.54 \\
    \cline{1-1}\cline{3-9}
        3 & & \multirow{3}{*}{\xmark} & FBK &  65.06 & 27.94 & 18.76 & 8.39 & 27.45 \\
        4 & & ~ & VR-SBE & 63.13 & 28.65 & 18,33 & 8.19 & 27.07 \\ 
        5 & & ~ & FBK+VR-SBE & 63.22 & 28.40 & 18.21 & 8.23 & 27.02 \\ 
    \cline{1-1}\cline{3-9}
        6 & & \multirow{3}{*}{VR-SBE} & FBK & 62.86 & 29.68 & 19.52 & 8.67 & 27.68 \\
        7 & & ~ & VR-SBE & 62.38 & 29.55 & 19.21 & 8.66 & 27.48 \\
        8 & & ~ & FBK+VR-SBE & \textbf{61.56} & \textbf{28.81} & 18.39 & 8.50 & \textbf{26.90} \\
    \hline\hline     
    \end{tabular}}
     \vspace{-3pt}
\end{table}

\begin{table}[htbp]
    \vspace{-3pt}
    \caption{Ablation study on the input features to the f-LHUC regression network on \textbf{DementiaBank Pitt.}}
    \label{tab:ablation-DBANK-hybrid}
    \centering
    \renewcommand\arraystretch{1.0}
    \renewcommand\tabcolsep{2.0pt}
    \scalebox{0.64}{\begin{tabular}{c|c|c|c|cc|cc|c}
    \hline\hline  
        \multirow{3}{*}{Sys.} & 
        \multirow{3}{*}{\tabincell{c}{Model\\(\#Para.)}} & 
        \multirow{3}{*}{\tabincell{c}{Adapt.\\Feat.}} & 
        \multirow{3}{*}{f-LHUC} &
        \multicolumn{5}{c}{WER\%} \\  
    \cline{5-9}
        & & & & \multicolumn{2}{c|}{Dev} & \multicolumn{2}{c|}{Eval} & \multirow{2}{*}{All (PAR)} \\
    \cline{5-8}
        & & & & INV & PAR & INV & PAR &  \\
    \hline\hline
        1 & \multirow{8}{*}{\tabincell{c}{Hybrid\\TDNN\\(18M)}} & \xmark & \multirow{2}{*}{\xmark} & 19.91 & 47.93 & 19.76 & 36.66 & 33.80 (44.59) \\
        2 & & VR-SBE & ~ & 18.72 & 44.67 & 18.65 & 34.03 & 31.55 (41.52) \\
    \cline{1-1}\cline{3-9}
        3 & & \multirow{3}{*}{\xmark} & FBK & 19.61 & 45.40 & 18.87 & 34.77 & 32.33 (42.25) \\
        4 & & ~ & VR-SBE & 19.60 & 46.91 & 18.09 & 34.98 & 32.94 (43.38) \\
        5 & & ~ & FBK+VR-SBE & 19.54 &  46.15 & 18.20 & 35.88 & 32.77 (43.11) \\
    \cline{1-1}\cline{3-9}
        6 & & \multirow{3}{*}{VR-SBE} & FBK & 19.95 & 45.70 & 19.20 & 34.41 & 32.54 (42.36) \\
        7 & & ~ & VR-SBE & 19.24 & 45.25 & 19.53 & 33.78 & 31.98 (41.85) \\
        8 & & ~ & FBK+VR-SBE & 17.87 & \textbf{43.83} & 16.87 & \textbf{34.56} & \textbf{30.91 (41.09)} \\
    \hline\hline     
    \end{tabular}}
     \vspace{-5pt}
\end{table}

\subsection{Dimensionality of F-LHUC Regression Targets}
\label{sec:f-LHUC_target_dim}


Similarly, when performing f-LHUC online speaker adaptation of Sec.~\ref{sec:f-LHUC}, Table~\ref{tab:ablation-UASpeech-f-LHUC-target} and~\ref{tab:ablation-DBANK-f-LHUC-target} suggest that setting the PCA projected speaker LHUC transform subspace dimensionality as 29 and 25 respectively for UASpeech and DementiaBank Pitt produce the best performance (Sys.6 vs. Sys.1-5 in Table~\ref{tab:ablation-UASpeech-f-LHUC-target},  Sys.1 vs. Sys.2-5 in Table~\ref{tab:ablation-DBANK-f-LHUC-target}) than other settings, for example, 5 on UASpeech intuitively representing the control and four dysarthric speech intelligibility based speakers grouping, or up to 400 for the DementiaBank Pitt which contains more speakers. These settings are used in all the following f-LHUC on-the-fly speaker adaptation experiments of this paper. 

\begin{table}[htbp]
    \vspace{-5pt}
    \caption{Ablation study on the f-LHUC regression network training targets dimensionality (Tar. Dim.) on \textbf{UASpeech}.}
    \label{tab:ablation-UASpeech-f-LHUC-target}
    \centering
    \renewcommand\arraystretch{1.0}
    \renewcommand\tabcolsep{2.0pt}
    \scalebox{0.69}{\begin{tabular}{c|c|c|c|c|cccc|c}
    \hline\hline  
        \multirow{2}{*}{Sys.} & 
        \multirow{2}{*}{\tabincell{c}{Model\\(\#Para.)}} & 
        \multirow{2}{*}{\tabincell{c}{Adapt.\\Feat.}} & 
        \multicolumn{2}{c|}{f-LHUC} &
        \multicolumn{5}{c}{WER\%} \\  
    \cline{4-10}
        & & & Inputs & Tar. Dim. & VL & L & M & H & All \\
    \hline\hline
        1 & \multirow{6}{*}{\tabincell{c}{Hybrid\\DNN\\(6M)}} & \multirow{6}{*}{VR-SBE} & \multirow{6}{*}{\tabincell{c}{FBK\\+\\VR-SBE}}  & 2 & 64.11 & 31.06 & 20.39 & 8.69 & 28.47 \\
        2 & & & & 5 & 63.81 & 30.51 & 20.19 & 8.59 & 28.19 \\
        3 & & & & 15 & 63.68 & 30.29 & 19.90 & 8.66 & 28.08 \\
        4 & & & & 20 & 62.58 & 29.60 & 19.00 & 8.70 & 27.51 \\
        5 & & & & 25 & 61.99 & 29.16 & 18.58 & 8.69 & 27.19 \\
        6 & & & & 29 & \textbf{61.56} & \textbf{28.81} & 18.39 & 8.50 & \textbf{26.90} \\
    \hline\hline     
    \end{tabular}}
     \vspace{-5pt}
\end{table}

\begin{table}[htbp]
    \vspace{-5pt}
    \caption{Ablation study on f-LHUC regression training targets dimensionality (Tar. Dim.) on \textbf{DementiaBank Pitt}.}
    \label{tab:ablation-DBANK-f-LHUC-target}
    \centering
    \renewcommand\arraystretch{1.0}
    \renewcommand\tabcolsep{2.0pt}
    \scalebox{0.64}{\begin{tabular}{c|c|c|c|c|cc|cc|c}
    \hline\hline  
        \multirow{3}{*}{Sys.} & 
        \multirow{3}{*}{\tabincell{c}{Model\\(\#Para.)}} & 
        \multirow{3}{*}{\tabincell{c}{Adapt.\\Feat.}} & 
        \multicolumn{2}{c|}{f-LHUC} &
        \multicolumn{5}{c}{WER\%} \\  
    \cline{4-10}
        & & & \multirow{2}{*}{Inputs} & \multirow{2}{*}{\tabincell{c}{Tar.\\Dim.}} & \multicolumn{2}{c|}{Dev} & \multicolumn{2}{c|}{Eval} & \multirow{2}{*}{All (PAR)} \\
    \cline{6-9}
        & & & & & INV & PAR & INV & PAR &  \\
    \hline\hline
        1 & \multirow{5}{*}{\tabincell{c}{Hybrid\\TDNN\\(18M)}} & \multirow{5}{*}{VR-SBE} & \multirow{5}{*}{\tabincell{c}{FBK\\+\\VR-SBE}} & 25 & 17.87 & \textbf{43.83} & 16.87 & \textbf{34.56} & \textbf{30.91 (41.09)} \\    
        2 & & & & 50 & 18.61 & 45.01 & 17.65 & 35.12 & 31.79 (42.08) \\    
        3 & & & & 100 & 18.64 & 44.66 & 17.76 & 34.60 & 31.35 (41.68) \\    
        4 & & & & 200 & 18.71 & 44.68 & 17.65 & 34.77 & 31.65 (41.75) \\    
        5 & & & & 400 & 18.54 & 44.64 & 17.31 & 35.23 & 31.62 (41.85) \\
    \hline\hline     
    \end{tabular}}
     \vspace{-10pt}
\end{table}

\subsection{History Interpolation Factor of F-LHUC Regression}
\label{secf:LHUC_interpolation_factor}



Tables~\ref{tab:ablation-UASpeech-f-LHUC-interpolation-factor} and~\ref{tab:ablation-DBANK-f-LHUC-interpolation-factor} suggest that setting the f-LHUC adaptation history interpolation factor (Eqn.~(\ref{eq:f-LHUC_calculation}), Sec.~\ref{sec:f-LHUC}) as $\alpha = 0.9$ produces the best performance (Sys.7 vs. others in Tables~\ref{tab:ablation-UASpeech-f-LHUC-interpolation-factor} and~\ref{tab:ablation-DBANK-f-LHUC-interpolation-factor}) on both UASpeech and DementiaBank Pitt. The DementiaBank Pitt task is more sensitive to $\alpha$, which may be attributed to its longer average sentence duration and temporal contexts than the UASpeech task (3.4s vs. 1.1s).

\begin{table}[htbp]
    \vspace{-5pt}
    \caption{Ablation study on history interpolation factor $\alpha$ (Int. Fac.) in f-LHUC regression training on \textbf{UASpeech}.}
    \label{tab:ablation-UASpeech-f-LHUC-interpolation-factor}
    \centering
    \renewcommand\arraystretch{1.0}
    \renewcommand\tabcolsep{2.0pt}
    \scalebox{0.69}{\begin{tabular}{c|c|c|c|c|cccc|c}
    \hline\hline  
        \multirow{2}{*}{Sys.} & 
        \multirow{2}{*}{\tabincell{c}{Model\\(\#Para.)}} & 
        \multirow{2}{*}{\tabincell{c}{Adapt.\\Feat.}} & 
        \multicolumn{2}{c|}{f-LHUC} &
        \multicolumn{5}{c}{WER\%} \\  
    \cline{4-10}
        & & & Inputs & Int. Fac. & VL & L & M & H & All \\
    \hline\hline
        1 & \multirow{8}{*}{\tabincell{c}{Hybrid\\DNN\\(6M)}} & \multirow{8}{*}{VR-SBE} & \multirow{8}{*}{\tabincell{c}{FBK\\+\\VR-SBE}} & 0.0 & 62.83 & 29.74 & 19.37 & 8.68 & 27.66 \\
        2 & & & & 0.1 & 63.47 & 29.70 & 19.31 & 8.71 & 27.78 \\
        3 & & & & 0.3 & 62.60 & 29.62 & 19.11 & 8.69 & 27.53 \\
        4 & & & & 0.5 & 62.58 & 29.71 & 19.03 & 8.71 & 27.54 \\
        5 & & & & 0.7 & 62.62 & 29.65 & 18.98 & 8.61 & 27.49 \\
        6 & & & & 0.8 & 62.35 & 29.61 & 19.19 & 8.70 & 27.50 \\
        7 & & & & 0.9 & \textbf{61.56} & \textbf{28.81} & 18.39 & 8.50 & \textbf{26.90} \\
        8 & & & & 1.0 & 62.29 & 29.58 & 19.13 & 8.64 & 27.45 \\
    \hline\hline
    \end{tabular}}
     \vspace{-10pt}
\end{table}












\begin{table}[htbp]
    \caption{Ablation study on history interpolation factor $\alpha$ in f-LHUC regression training on \textbf{DementiaBank Pitt}.}
    \label{tab:ablation-DBANK-f-LHUC-interpolation-factor}
    \centering
    \renewcommand\arraystretch{1.0}
    \renewcommand\tabcolsep{2.0pt}
    \scalebox{0.64}{\begin{tabular}{c|c|c|c|c|cc|cc|c}
    \hline\hline  
        \multirow{3}{*}{Sys.} & 
        \multirow{3}{*}{\tabincell{c}{Model\\(\#Para.)}} & 
        \multirow{3}{*}{\tabincell{c}{Adapt.\\Feat.}} & 
        \multicolumn{2}{c|}{f-LHUC} &
        \multicolumn{5}{c}{WER\%} \\  
    \cline{4-10}
        & & & \multirow{2}{*}{Inputs} & \multirow{2}{*}{\tabincell{c}{Int.\\Fac.}} & \multicolumn{2}{c|}{Dev} & \multicolumn{2}{c|}{Eval} & \multirow{2}{*}{All (PAR)} \\
    \cline{6-9}
        & & & & & INV & PAR & INV & PAR &  \\
    \hline\hline
        1 & \multirow{8}{*}{\tabincell{c}{Hybrid\\TDNN\\(18M)}} & \multirow{8}{*}{VR-SBE} & \multirow{8}{*}{\tabincell{c}{FBK\\+\\VR-SBE}} & 0.0 & 19.44 & 45.40 & 17.65 & 35.04 & 32.27 (42.33) \\
        2 & & & & 0.1 & 19.57 & 44.88 & 18.53 & 33.84 & 31.95 (41.61) \\    
        3 & & & & 0.3 & 19.48 & 45.21 & 18.20 & 33.95 & 32.05 (41.88) \\    
        4 & & & & 0.5 & 19.89 & 45.12 & 19.09 & 34.10 & 32.23 (41.86) \\    
        5 & & & & 0.7 & 19.69 & 45.05 & 18.65 & 34.31 & 32.14 (41.87) \\
        6 & & & & 0.8 & 19.13 & 44.45 & 17.98 & 34.24 & 31.65 (41.43) \\
        7 & & & & 0.9 & 17.87 & \textbf{43.83} & 16.87 & \textbf{34.56} & \textbf{30.91 (41.09)} \\
        8 & & & & 1.0 & 18.81 & 44.98 & 17.54 & 34.73 & 31.80 (41.95) \\    
    \hline\hline     
    \end{tabular}}
     \vspace{-10pt}
\end{table}

\vspace{-0.5em}
\subsection{Baseline Adaptation of E2E Conformer Models }
\label{sec:LHUC_in_e2e}


\begin{table}[ht]
    \vspace{0pt}
    \caption{Ablation study on performance comparison between feature-based and supervised LHUC-based speaker adaptation of Conformer (CONF.) models on \textbf{UASpeech}. ``conv2d'' and ``enc-x'' denote the convolution subsampling module and the x-th encoder block. }
    \label{tab:ablation-UASpeech-conformer}
    \Large
    \centering
    \renewcommand\arraystretch{1.0}
    \renewcommand\tabcolsep{2.0pt}
    \scalebox{0.50}{\begin{tabular}{c|c|c|c|c|c|cccc|c}
    \hline\hline  
        \multirow{2}{*}{Sys.} & 
        \multirow{2}{*}{\tabincell{c}{Model\\(\#Para.)}} & 
        \multirow{2}{*}{\tabincell{c}{Data\\Aug.}} &
        \multirow{2}{*}{\#Hrs} &
        \multirow{2}{*}{\tabincell{c}{Adapt.\\Feat.}}  &
        \multirow{2}{*}{\tabincell{c}{LHUC\\SAT}} & 
        \multicolumn{5}{c}{WER\%} \\  
    \cline{6-11}
        & & & & & & VL & L & M & H & All \\
    \hline\hline
        1 & \multirow{8}{*}{\tabincell{c}{CONF.\\(52M)}} & \multirow{8}{*}{\cmark} & \multirow{8}{*}{173} & \multirow{5}{*}{\xmark} & \xmark & 65.70 & 40.63 & 33.39 & 9.53 & 34.07 \\
    \cline{1-1}\cline{6-11}
        2 & & & & & conv2d & 64.79 & 39.01 & 32.68 & 11.11 & 33.86 \\
        3 & & & & & enc-1 & 66.54 & 41.77 & 32.98 & 11.03 & 34.97 \\
        4 & & & & & enc-6 & 65.40 & 43.10 & 33.49 & 11.56 & 35.35 \\
        5 & & & & & enc-12 & 65.59 & 42.29 & 34.45 & 12.85 & 35.80 \\
    \cline{1-1}\cline{5-11}
        6 & & & & iVector & \multirow{3}{*}{\xmark} & 69.05 & 42.45 & 33.60 & 9.74 & 35.37 \\
        7 & & & & xVector & & 67.70 & 40.13 & 30.80 & 8.49 & 33.52 \\
        8 & & & & VR-SBE & & \textbf{67.52} & \textbf{38.85} & 28.60 & 7.88 & \textbf{32.52}  \\
    \hline\hline     
    \end{tabular}}
     \vspace{-4pt}
\end{table}

\begin{table}[ht]
    \vspace{-4pt}
    \caption{Ablation study on performance comparison between feature-based and supervised LHUC speaker adaptation of Conformer (CONF.) models on \textbf{DementiaBank Pitt}.}
    \label{tab:ablation-DBANK-conformer}
    \Large
    \centering
    \renewcommand\arraystretch{1.0}
    \renewcommand\tabcolsep{2.0pt}
    \scalebox{0.44}{\begin{tabular}{c|c|c|c|c|c|cc|cc|c}
    \hline\hline  
        \multirow{3}{*}{Sys.} & 
        \multirow{3}{*}{\tabincell{c}{Model\\(\#Para.)}} & 
        \multirow{3}{*}{\tabincell{c}{Data\\Aug.}} &
        \multirow{3}{*}{\#Hrs} &
        \multirow{3}{*}{\tabincell{c}{Adapt.\\Feat.}}  &
        \multirow{3}{*}{\tabincell{c}{LHUC\\SAT}} & 
        \multicolumn{5}{c}{WER\%} \\  
    \cline{7-11}
        & & & & & & \multicolumn{2}{c|}{Dev} & \multicolumn{2}{c|}{Eval} & \multirow{2}{*}{All (PAR)} \\
    \cline{7-10}
        & & & & & & INV & PAR & INV & PAR &  \\
    \hline\hline
        1 & \multirow{8}{*}{\tabincell{c}{CONF.\\(52M)}} & \multirow{8}{*}{\cmark} & \multirow{8}{*}{58.9} & \multirow{5}{*}{\xmark} & \xmark & 20.97 & 48.71 & 19.42 & 36.93 & 34.57 (45.22) \\
    \cline{1-1}\cline{6-11}
        2 & & & & & conv2d & 21.44 & 47.39 & 19.53 & 37.16 & 34.28 (44.36) \\
        3 & & & & & enc-1 & 22.51 & 50.56 & 21.86 & 40.85 & 36.66 (47.69) \\
        4 & & & & & enc-6 & 22.66 & 50.16 & 20.19 & 39.63 & 36.31 (47.04) \\
        5 & & & & & enc-12 & 22.74 & 50.12 & 20.64 & 39.68 & 36.35 (47.03) \\
    \cline{1-1}\cline{5-11}
        6 & & & & iVector & \multirow{3}{*}{\xmark} & 21.48 & 48.32 & 17.42 & 37.79 & 34.71 (45.20) \\
        7 & & & & xVector & & 21.77 & 49.38 & 18.09 & 37.83 & 35.27 (45.96) \\
        8 & & & & VR-SBE & & 20.83 & \textbf{47.39} & 17.64 & \textbf{36.34} & \textbf{33.84 (44.12)} \\
    \hline\hline     
    \end{tabular}}
     \vspace{-5pt}
\end{table}

Baseline adaptation of E2E Conformer models of this paper uses feature-based on-the-fly adaptation approaches only that are based on, e.g. iVectors or xVectors. Such a choice is based on the ablation studies in Tables~\ref{tab:ablation-UASpeech-conformer} and~\ref{tab:ablation-DBANK-conformer} on UASpeech\renewcommand\footnoterule{\vspace{-8pt}}\footnote{Block 2 data of the control speech and its speed-perturbed
versions are also used for Conformer system training.} and DementiaBank Pitt. The performance of iVector, xVector and VR-SBE adaptation (Sys.6-8) are found generally more competitive than those obtained using LHUC adaptation performed in supervised mode at different Conformer sublayers (Sys.2-5), serving as the upper bound performance of model-based adaptation approaches including both LHUC and f-LHUC.

\section{Experiments and Results}
\label{sec:exp}

In this section, we investigate the performance\renewcommand\footnoterule{\vspace{-8pt}}\footnote{For all speech recognition results measured in WER/CER, a matched pairs sentence-segment word error (MAPSSWE) statistical signiﬁcance test~\cite{gillick1989some} is performed at the signiﬁcance level $\alpha=0.05$.} of our feature-based on-the-fly adaptation approaches on four tasks: the English UASpeech~\cite{kim2008dysarthric} and TORGO~\cite{rudzicz2012torgo} dysarthric speech datasets, and the English DementiaBank Pitt~\cite{becker1994natural} and Cantonese JCCOCC MoCA~\cite{xu2021speaker} elderly speech corpora. All the settings determined in the ablation studies of Sec.~\ref{sec:implement} are adopted. SI and SD speed perturbations ~\cite{geng2020investigation,ye2021development} based data augmentation are applied to all tasks. The extraction of 100-dim iVector\renewcommand\footnoterule{\vspace{-8pt}}\footnote{Kaldi: egs/wsj/s5/local/nnet3/run\_ivector\_common.sh} and 25-dim xVector\renewcommand\footnoterule{\vspace{-8pt}}\footnote{Kaldi: egs/sre16/v1/local/nnet3/xvector/tuning/run\_xvector\_1a.sh} follow the Kaldi recipes using the same training data as for ASR systems\renewcommand\footnoterule{\vspace{-8pt}}\footnote{Changing the dimensionality of iVector and xVector to other settings produced marginal and statistically non-significant differences in performance.}.

\vspace{-0.5em}
\subsection{Experiments on Dysarthric Speech}
\label{sec:exp_dys}

\subsubsection{The UASpeech Dataset}
\label{sec:exp_dys_ua_intro}

As the largest publicly available dysarthric speech dataset, UASpeech~\cite{kim2008dysarthric} is an isolated word recognition task in English with 103 hours of speech from 16 dysarthric and 13 control speakers recorded using a 7-channel microphone array. The block based training-evaluation data partitioning protocol~\cite{kim2008dysarthric,christensen2012comparative,sehgal2015model,hu2019cuhk,christensen2013combining} is adopted. For each speaker, the data is split into B1, B2 and B3, each with the same 155 common words and a different 100 uncommon words. The training set combines B1 and B3 data of all 29 speakers, while the evaluation set contains the B2 data from the 16 dysarthric speakers. Silence stripping~\cite{liu2021recent} leads to a 30.6h training set (99195 utt.) and a 9h evaluation set (26520 utt.). Further data augmentation~\cite{geng2020investigation,liu2021recent} produces a 130.1h training set (399110 utt.). As E2E systems are sensitive to training data coverage, B2 of the control speech and its speed-perturbed
versions are also used for Conformer system training, creating a 173h training set (538292 utt.).

\subsubsection{The TORGO Dataset}
\label{sec:exp_dys_torgo_intro}

The English TORGO~\cite{rudzicz2012torgo} dysarthric dataset contains 13.5h of speech from 8 dysarthric and 7 control speakers based on short sentences and single words. We adopt a 3-block based training-evaluation data partition similar to UASpeech. The control speech and two-thirds of the speech of each impaired speaker are used for training, while the remaining one-third is for evaluation. Silence stripping produces a 6.5h training set (14541 utt.) and a 1h evaluation set (1892 utt.). Further speed perturbation~\cite{geng2020investigation,hu2022exploit}  produces a 34.1h augmented training set (61813 utt.).

\subsubsection{Experimental Setup for UASpeech}
\label{sec:exp_dys_ua_setup}

Following~\cite{geng2020investigation,liu2021recent}, the hybrid DNN systems with six $2000$-dim and one $100$-dim layers are implemented using extended Kaldi~\cite{povey2011kaldi}. Each hidden layer contains linear bottleneck projection, affine transformation, rectified linear unit (ReLU) activation, batch normalization and dropout operations, while Softmax activation is used in the output layer. Two skipping connections respectively feed the first layer’s outputs to the third and those of the fourth to the sixth. Multi-task learning is adopted, using frame-level triphone state and mono-phone alignments as targets. The E2E Conformers are built using ESPnet~\cite{watanabe2018espnet}\renewcommand\footnoterule{\vspace{-8pt}}\footnote{12 encoder + 12 decoder layers, feed-forward layer dim = 2048, attention heads = 4, dim of attention heads = 256, interpolated CTC+AED cost.} to predict grapheme (letter) sequence outputs. The inputs to both systems are 80-dim FBK + $\Delta$ features, while a 9-frame context is used in hybrid DNN. A uniform language model with a word grammar network is adopted during evaluation~\cite{christensen2012comparative}.

\subsubsection{Experimental Setup for TORGO}
\label{sec:exp_dys_torgo_setup}

The hybrid factorized TDNN systems with 7 context slicing layers are constructed following Kaldi~\cite{povey2011kaldi} chain recipe. The setup of E2E graphemic Conformers is similar to that for UASpeech\renewcommand\footnoterule{\vspace{-8pt}}\footnote{8 encoder + 4 decoder layers, feed-forward layer dim = 1024}. 40-dim FBK features are used as input for both systems, with a 3-frame context for the hybrid system. A 3-gram language model (LM) trained using all TORGO transcripts with a word recognition vocabulary of 1.6k is utilized during evaluation.

\subsubsection{Performance Analyses}
\label{sec:exp_dys_analysis}

\begin{table}[ht]
    \vspace{0pt}
    \caption{Performance comparison between baseline iVector, xVector, batch-mode LHUC or SBE~\cite{geng2022spectro} speaker adaptation and the proposed VR-SBE on-the-fly speaker adaptation on the \textbf{UASpeech} test set of 16 dysarthric speakers. ``CONF.'' denotes Conformer. ``VL'', ``L'', ``M'' and ``H'' refer to speech intelligibility ``very low'', ``low'', ``mid'' and ``high''. ``On Fly'' indicates using on-the-fly adaptation. $^\dag$, $^\ddag$ and $^\star$ denote a stat. significant improvement ($\alpha=0.05$) obtained over iVector (Sys.2,10,14,22,27), xVector (Sys.3,11,15,23,28), or both. ``A $\rightarrow$ B'' denotes system A produces N-best outputs in the $1^{st}$ decoding pass before the $2^{nd}$ pass rescoring by system B.}
    \label{tab:recog-UASpeech}
    \centering
    \renewcommand\arraystretch{1.0}
    \renewcommand\tabcolsep{2.0pt}
    \scalebox{0.63}{\begin{tabular}{c|c|c|c|c|c|c|cccc|c}
    \hline\hline  
        \multirow{2}{*}{Sys.} & 
        \multirow{2}{*}{\tabincell{c}{Model\\(\#Para.)}} & 
        \multirow{2}{*}{\#Hrs} &
        \multirow{2}{*}{\tabincell{c}{Adapt.\\Feat.}} & 
        \multirow{2}{*}{\tabincell{c}{LHUC\\SAT}} &
        \multirow{2}{*}{f-LHUC} &
        \multirow{2}{*}{\tabincell{c}{On\\Fly}} &
        \multicolumn{5}{c}{WER\%} \\
    \cline{8-12}
       & & & & & & & VL & L & M & H & All \\
    \hline\hline
        1 & \multirow{12}{*}{\tabincell{c}{Hybrid\\DNN\\(6M)}} & \multirow{12}{*}{30.6} & \xmark & \multirow{6}{*}{\xmark} & \multirow{6}{*}{\xmark} & - & 69.82 & 32.61 & 24.53 & 10.40 & 31.45 \\
        2 & & & iVector & & & \cmark & 69.46 & 33.78 & 22.58 & 10.45 & 31.33 \\
        3 & & & xVector & & & \cmark & 68.07 & 33.56 & 24.07 & 9.49 & 30.93 \\
        4 & & & SBE~\cite{geng2021spectro} & & & \xmark & 64.43 & 29.71 & 19.84 & 8.57 & 28.05 \\
        5 & & & iVR & & & \cmark & 68.66 & 33.72 & 22.84 & 9.83 & 30.99 \\
        6 & & & VR-SBE & & & \cmark & 65.04$^\star$ & 30.90$^\star$ & 20.70$^\star$ & 10.15$^\star$ & \textbf{28.85$^\star$} \\
    \cline{1-1}\cline{4-12}
        7 & & & \xmark & \multirow{2}{*}{\cmark} & \multirow{2}{*}{\xmark} & \xmark & 64.39 & 29.88 & 20.27 & 8.95 & 28.29 \\
        8 & & & SBE~\cite{geng2021spectro} & & & \xmark & 63.40 & 28.90 & 18.64 & 8.13 & 27.24 \\
    \cline{1-1}\cline{4-12}
        9 & & & \xmark & \multirow{3}{*}{\xmark} & (FBK) & \cmark & 66.47$^\dag$ & 29.55$^\dag$ & 21.00$^\dag$ & 8.99$^\dag$ & 28.80$^\dag$ \\
        10 & & & iVector & & (+iVector) & \cmark & 64.86 & 36.44 & 21.17 & 9.03 & 30.29 \\
        11 & & & xVector & & (+xVector) & \cmark & 68.28 & 30.79 & 21.98 & 9.31 & 29.80 \\
        12 & & & VR-SBE & & (+VR-SBE) & \cmark & 65.75$^\ddag$ & 29.80$^\star$ & 19.07$^\star$ & 8.99$^\ddag$ & \textbf{28.31$^\star$} \\
    \cline{1-1}\cline{4-12}
        6+12 & & & \multicolumn{3}{c|}{-} & \cmark & \textbf{64.36$^\star$} & \textbf{29.68$^\star$} & 
        18.96$^\star$ & 
        8.89$^\star$ & 
        \textbf{27.96$^\star$} \\
    \hline\hline
        13 & \multirow{12}{*}{\tabincell{c}{Hybrid\\DNN\\(6M)}} & \multirow{12}{*}{130.1} & \xmark & \multirow{6}{*}{\xmark} & \multirow{6}{*}{\xmark} & - & 66.45 & 28.95 & 20.37 & 9.62 & 28.73 \\
        14 & & & iVector & & & \cmark & 65.73 & 30.10 & 20.21 & 9.03 & 28.65 \\
        15 & & & xVector & & & \cmark & 65.40 & 29.32 & 20.72 & 8.85 & 28.41 \\
        16 & & & SBE~\cite{geng2021spectro} & & & \xmark & 61.55 & 27.52 & 17.31 & 8.22 & 26.26 \\
        17 & & & iVR & & & \cmark & 66.02 & 29.52 & 19.56 & 9.32 & 28.53 \\
        18 & & & VR-SBE & & & \cmark & 62.54$^\star$ & 30.22$^\ddag$ & 18.54$^\star$ & 8.59$^\star$ & \textbf{27.54$^\star$} \\
    \cline{1-1}\cline{4-12}
        19 & & & \xmark & \multirow{2}{*}{\cmark} & \multirow{2}{*}{\xmark} & \xmark & 62.50 & 27.26 & 18.41 & 8.04 & 26.55 \\
        20 & & & SBE~\cite{geng2021spectro} & & & \xmark & 59.83 & 27.16 & 16.80 & 7.91 & 25.60 \\
    \cline{1-1}\cline{4-12}
        21 & & & \xmark & \multirow{3}{*}{\xmark} & (FBK) & \cmark & 65.06$^\dag$ & 27.94$^\dag$ & 18.76$^\dag$ & 8.39$^\dag$ & 27.45$^\dag$ \\
        22 & & & iVector & & (+iVector) & \cmark & 63.63 & 32.56 & 18.52 & 8.31 & 28.28 \\
        23 & & & xVector & & (+xVector) & \cmark & 65.70 & 29.22 & 20.13 & 8.90 & 28.35 \\
        24 & & & VR-SBE & & (+VR-SBE) & \cmark & 61.56$^\star$ & 28.81$^\star$ & 18.39$^\ddag$ & 8.50$^\ddag$ & \textbf{26.90$^\star$} \\
    \cline{1-1}\cline{4-12}
        18+24 & & & \multicolumn{3}{c|}{-} & \cmark &\textbf{60.80$^\star$} &\textbf{28.19$^\star$}
        & 17.72$^\star$  
        & 8.23$^\star$
        & \textbf{26.36$^\star$} \\
    \hline\hline
        25 & \multirow{7}{*}{\tabincell{c}{CONF.\\(52M)}} & 130.1 & \xmark & \multirow{7}{*}{\xmark} & \multirow{7}{*}{\xmark} & - & 73.88 & 53.12 & 49.92 & 42.03 & 53.17 \\
    \cline{3-4}\cline{7-12}
        26 & & \multirow{6}{*}{173} & \xmark & & & - & 65.70 & 40.63 & 33.39 & 9.53 & 34.07 \\
        27 & & & iVector & & & \cmark & 69.05 &  42.45 & 33.60 & 9.74 & 35.37 \\
        28 & & & xVector & & & \cmark & 67.70 & 40.13 & 30.80 & 8.49 & 33.52 \\
        29 & & & SBE~\cite{geng2021spectro} & & & \xmark & 65.18 & 34.90 & 24.21 & 5.00 & 29.19 \\
        30 & & & iVR & & & \cmark & 68.94 & 42.00 & 32.19 & 8.52 & 34.55 \\
        31 & & & VR-SBE & & & \cmark & \textbf{67.52$^\dag$} & \textbf{38.85$^\star$} 
        & 28.60$^\star$ 
        & 7.88$^\star$
        & \textbf{32.52$^\star$} \\
    \hline\hline
        32 & \multicolumn{6}{c|}{18+24 $\rightarrow$ 31} & \textbf{57.33$^\star$} &\textbf{24.79$^\star$} 
        & 14.27$^\star$ 
        & 5.99$^\star$
        & \textbf{23.33$^\star$} \\
    \hline\hline
    \end{tabular}}
    \vspace{-5pt}
\end{table}

The performance contrast between baseline iVector, xVector, batch-mode LHUC or SBE speaker adaptation, and the proposed VR-SBE on-the-fly speaker adaptation on the UASpeech task is shown in Table~\ref{tab:recog-UASpeech}. Several trends can be observed: 

i) Our proposed on-the-fly VR-SBE adapted systems (Sys.6,18,31) consistently outperform iVector (Sys.2,14,27) and xVector adaptation (Sys.3,15,28) with varying amounts of training data on both hybrid DNN and E2E Conformer systems. Statistically significant overall word error rate (WER) reductions by up to \textbf{2.48\% abs. (7.92\% rel.) and 2.85\% abs. (8.06\% rel.)} can be respectively achieved on DNN (Sys.6 vs. Sys.2) and Conformer (Sys.31 vs. Sys.27) models. 

ii) Our on-the-fly VR-SBE adaptation (Sys.6,18,31) consistently outperforms the comparable variance-regularized iVector (iVR) adaptation (Sys.5,17,30). 

iii) The improvements obtained by batch-mode LHUC adaptation (Sys.7,19) over the SI systems (Sys.1,13) are largely retained (by 82\%) and comparable to those obtained using on-the-fly VR-SBE adaptation (Sys.6,18). 

iv) On-the-fly VR-SBE adaptation (Sys.6,18) also produces performance comparable to batch-mode SBE adaptation~\cite{geng2022spectro} that requires speaker-level embedding averaging (Sys.4,16) and incurs additional processing latency. 

v) The combined use of both input speaker features and f-LHUC adaptation leads to the most powerful form of on-the-fly adaptation configurations (Sys.10-12, Sys.22-24). Among these, on-the-fly VR-SBE plus FBK+VR-SBE driven f-LHUC (Sys.12,24) adaptation not only consistently outperforms the comparable baselines by replacing VR-SBE with iVector or xVector (Sys.10-11, Sys.22-23), but also produces further WER reductions over using VR-SBE adaptation alone by up to 0.64\% abs. (2.32\% rel.) (Sys.12,24 vs. Sys.6,18). 

vi) The best performance is obtained by combining the system using VR-SBE plus FBK+VR-SBE driven f-LHUC adaptation (Sys.24) and using only VR-SBE adaptation alone (Sys.18) via frame-level joint decoding (Sys.18+24)\renewcommand\footnoterule{\vspace{-8pt}}\footnote{System weights empirically set as 11:9.}, before the resulting 50-best outputs are further rescored by the VR-SBE adapted Conformer\renewcommand\footnoterule{\vspace{-8pt}}\footnote{$1^{st}$ and $2^{nd}$ pass system weights empirically set as 1:9.}. Using this combined system (Sys.32), statistically significant overall WER reductions by up to \textbf{5.32\% abs. (18.57\% rel.)} can be achieved over iVector and xVector adaptation (Sys.32 vs. Sys.14).

Performance comparison between our best on-the-fly speaker-adapted system (Sys.32, Table~\ref{tab:recog-UASpeech}) and recently published systems on UASpeech is presented in Table~\ref{tab:UA-compare}. To our best knowledge, our system (Sys.32, Table~\ref{tab:recog-UASpeech}) gives the best performance among all systems in Table~\ref{tab:UA-compare} without using out-of-domain data and self-supervised learning (SSL). It produces performance comparable or superior to the fine-tuned SSL systems (cross-lingual XLRS, WavLM, or Wav2vec2.0), particularly on the most challenging ``VL'' intelligibility subset. 

\begin{table}[ht]
    \vspace{-5pt}
  \caption{Performance against recently published systems on \textbf{UASpeech}. ``DA'' denotes data augmentation. ``SSL'' refers to incorporating fine-tuned self-supervised learning foundation speech models or features.}
  \label{tab:UA-compare}
  \Large
  \centering
  \vspace{-0.5em}
  \renewcommand\arraystretch{1}
  \scalebox{0.34}{\begin{tabular}{ccccc}
  \toprule
    Sys. & Online Adapt & SSL & VL & All \\
  \midrule
    Sheffield-2015 speaker adaptive training~\cite{sehgal2015model} & \xmark & \xmark & 70.78 & 34.85 \\
    Sheffield-2020 fine-tuning CNN-TDNN speaker adaptation~\cite{xiong2020source} & \cmark & \xmark & 68.24 & 30.76 \\
    CUHK-2020 DNN + DA + LHUC SAT~\cite{geng2020investigation} & \xmark & \xmark & 62.44 & 26.37 \\ 
    CUHK-2021 QuartzNet + CTC + meta-learning + SAT~\cite{wang2021improved} & \xmark & \xmark & 69.30 & 30.50 \\
    Sheffield-2022 DA + source filter features + iVector adapt~\cite{yue2022acoustic} & \cmark & \xmark & - & 30.30 \\
    FAU-2022 cross-lingual XLRS + Conformer \cite{hernandez22_interspeech} & -  & \cmark & 62.00 & 26.10 \\
    Nagoya-2022 WavLM~\cite{violeta2022investigating} & - & \cmark & 71.50 & 51.80 \\
    BUT-2022 Wav2vec2.0 + fMLLR + xvectors~\cite{baskar2022speaker} & \cmark & \cmark & 57.72 & 22.83  \\
    JHU-2023 DuTa-VC (Diffusion) + Conformer~\cite{wang23qa_interspeech} & - & \xmark & 63.70 & 27.90\\
    \textbf{DA + SVR adapt + f-LHUC system combination + 2-pass rescoring (Table~\ref{tab:recog-UASpeech}, Sys.32)} & \cmark & \xmark & \textbf{57.33} & \textbf{23.33} \\
  \bottomrule
  \end{tabular}}
\end{table}

\begin{table}[htbp]
    \caption{Performance comparison between baseline iVector, xVector, batch-mode LHUC or SBE~\cite{geng2022spectro} adaptation and the proposed VR-SBE on-the-fly speaker adaptation on \textbf{TORGO}. ``Seve.'', ``Mod.'' and ``Mild'' refer to the speech intelligibility. $^\dag$ and $^\star$ denote a stat. significant improvement ($\alpha=0.05$) obtained over iVector adaptation (Sys.2,10,14), or further over xVector adaptation (Sys.3,11,15).}
    \label{tab:recog-TORGO}
    \centering
    \renewcommand\arraystretch{1.0}
    \renewcommand\tabcolsep{2.0pt}
      \scalebox{0.78}{\begin{tabular}{c|c|c|c|c|c|ccc|c}
    \hline\hline  
        \multirow{2}{*}{Sys.} & 
        \multirow{2}{*}{\tabincell{c}{Model\\(\#Para.)}} & 
        \multirow{2}{*}{\tabincell{c}{Adapt.\\Feat.}} & 
        \multirow{2}{*}{\tabincell{c}{LHUC}} &
        \multirow{2}{*}{f-LHUC} &
        \multirow{2}{*}{\tabincell{c}{On\\Fly}} &
        \multicolumn{4}{c}{WER\%} \\
    \cline{7-10}
       & & & & & & Seve. & Mod. & Mild & All \\
    \hline\hline
        1 & \multirow{12}{*}{\tabincell{c}{Hybrid\\TDNN\\(10M)}} & \xmark & \multirow{6}{*}{\xmark} & \multirow{6}{*}{\xmark} & - & 12.80 & 8.78 & 3.64 & 9.47 \\
        2 & & iVector & & & \cmark & 13.82 & 5.92 & 2.40 & 9.07 \\
        3 & & xVector & & & \cmark & 13.86 & 4.39 & 3.02 & 8.94 \\
        4 & & SBE~\cite{geng2022spectro} & & & \xmark & 11.67 & 4.59 & 2.86 & 7.80 \\
        5 & & iVR & & & \cmark & 13.62 & 5.31 & 2.86 & 8.96  \\
        6 & & VR-SBE & & & \cmark & 12.07$^\star$ & 4.69$^\dag$ & 2.63 & \textbf{7.97$^\star$} \\
    \cline{1-1}\cline{3-10}
        7 & & \xmark & \multirow{2}{*}{\cmark} & \multirow{2}{*}{\xmark} & \xmark & 12.60 & 8.78 & 3.64 & 9.36 \\
        8 & & SBE~\cite{geng2022spectro} & & & \xmark & 11.71 & 4.29 & 2.86 & 7.76 \\
    \cline{1-1}\cline{3-10}
        9 & & \xmark & \multirow{3}{*}{\xmark} & (FBK) & \cmark & 12.80 & 5.00 & 2.94 & 8.50 \\
        10 & & iVector & & (+iVector) & \cmark & 13.94 & 5.61 & 2.55 & 9.11 \\
        11 & & xVector & & (+xVector) & \cmark & 14.35 & 5.82 & 2.55 & 9.36  \\
        12 & & VR-SBE & & (+VR-SBE) & \cmark & 12.07$^\star$ & 4.90$^\star$ & 2.79 & \textbf{8.05$^\star$} \\
    \cline{1-1}\cline{3-10}
        6+12 & & \multicolumn{3}{c|}{-} & \cmark & \textbf{11.59$^\star$} & \textbf{4.69$^\star$} & 2.71 & \textbf{7.73$^\star$} \\ 
    \hline\hline
        13 & \multirow{6}{*}{\tabincell{c}{CONF.\\(18M)}} & - & \multirow{6}{*}{\xmark} & \multirow{6}{*}{\xmark} & - & 21.66 & 6.22 & 4.10 & 13.67 \\
        14 & & iVector & & & \cmark & 22.15 & 7.44 & 3.94 & 14.13 \\
        15 & & xVector & & & \cmark & 21.38 & 9.08 & 4.02 & 14.09 \\
        16 & & SBE~\cite{geng2022spectro} & & & \xmark & 18.69 & 6.83 & 3.71 & 12.15 \\
        17 & & iVR & & & \cmark & 21.34 & 7.65 & 3.63 & 13.67 \\
        18 & & VR-SBE & & & \cmark & \textbf{19.22$^\star$} & \textbf{6.32$^\star$} & 3.94 & \textbf{12.38$^\star$} \\
    \hline\hline
    \end{tabular}}
    \vspace{-5pt}
\end{table}

Similar trends can be observed on a comparable set of experiments conducted on TORGO~\cite{rudzicz2012torgo} (Table~\ref{tab:recog-TORGO}). Our on-the-fly VR-SBE adaptation (Sys.6,18) outperforms iVector (Sys.2,14) or xVector (Sys.3,15) adaptation by up to \textbf{1.75\% abs. (12.38\% rel.)} WER reduction (Sys.18 vs. Sys.14), while outperforming batch-mode LHUC adaptation by \textbf{1.39\% abs. (14.85\% rel.)} WER reduction (Sys.6 vs. Sys.7). Combining on-the-fly VR-SBE and FBK+VR-SBE driven f-LHUC adaptation leads to up to \textbf{1.34\% abs.(14.77\% rel.)} WER reduction over iVector and xVector adaptation (Sys.6+12 vs. Sys.2).

\subsubsection{Processing Latency Analyses}
\label{sec:exp_dys_latency}

We then conduct further analysis on the processing latency and feature homogeneity of our on-the-fly adaptation techniques. The size of the analysis sliding window imposed on the VR-SBE feature extraction is gradually reduced to as short as 10 ms of acoustic inputs.  

\begin{table}[htbp]
\vspace{0pt}
    \caption{Ablation study of the analysis sliding window (Slid. Wind.) size for on-the-fly VR-SBE feature extraction on \textbf{UASpeech}. ``RTF'' denotes real-time factor.  $^\dag$ and $^\star$ denote a stat. significant improvement ($\alpha=0.05$) obtained over iVector (Sys.1), or both iVector and xVector (Sys.1-2).}
    \label{tab:VR-SBE-sliding-window-UASpeech}
    \centering
    \renewcommand\arraystretch{1.0}
    \renewcommand\tabcolsep{2.0pt}
    \scalebox{0.75}{\begin{tabular}{c|c|c|c|c|cccc|c}
    \hline\hline  
        \multirow{2}{*}{Sys.} & 
        \multirow{2}{*}{\tabincell{c}{Model\\(\#Para.)}} & 
        \multirow{2}{*}{\tabincell{c}{Adapt.\\Feat.}} & 
        \multirow{2}{*}{\tabincell{c}{Slid.\\Wind.}} &
        \multirow{2}{*}{RTF} &
        \multicolumn{5}{c}{WER\%} \\
    \cline{6-10}
         & & & & & VL & L & M & H & All \\
    \hline\hline
        1 & \multirow{10}{*}{\tabincell{c}{Hybrid\\DNN\\(6M)}} & iVector & 100ms & 0.10 & 65.73 & 30.10 & 20.21 & 9.03 & 28.65 \\
        2 & ~ & xVector & utt. & 1.01 & 65.40 & 29.32 & 20.72 & 8.85 & 28.41 \\
    \cline{1-1} \cline{3-10}
        3 & & \multirow{8}{*}{VR-SBE} & utt. & 1.02 & 62.54$^\star$ & 30.22 & 18.54$^\dag$ & 8.59$^\dag$ & 27.54$^\star$ \\
        4 & & & 250ms & 0.23 & 64.26$^\star$ & 28.62$^\star$ & 19.90$^\star$ & 9.04 & 27.89$^\star$ \\ 
        5 & & & 150ms & 0.14 & 64.18$^\star$ & 28.71$^\star$ & 19.82$^\star$ & 9.29 & 27.97$^\star$ \\
        6 & & & 100ms & 0.10 & 64.49$^\star$ & 28.36$^\star$ & 19.47$^\star$ & 9.27 & 27.87$^\star$ \\
        7 & & & 50ms & 0.06 & 63.99$^\star$ & 28.68$^\star$ & 19.17$^\star$ & 8.99 & 27.70$^\star$ \\
        8 & & & 30ms & 0.04 & 64.20$^\star$ & 29.29$^\dag$ & 19.39$^\star$ & 9.21 & 28.02$^\star$ \\
        9 & & & 20ms & 0.03 & 64.38$^\star$ & 28.99$^\dag$ & 19.47$^\star$ & 9.19 & 27.98$^\star$ \\
        10 & & & 10ms & 0.03 & \textbf{64.77$^\star$} & \textbf{29.03$^\dag$} & \textbf{19.21$^\star$} & 9.09 & \textbf{28.00$^\star$} \\
    \hline\hline 
    \end{tabular}}
\vspace{-5pt}
\end{table}

As shown in Table~\ref{tab:VR-SBE-sliding-window-UASpeech}, the performance of our on-the-fly VR-SBE adaptation is largely insensitive to the duration of analysis sliding windows from 1 utt. to 10 ms (Sys.3-10). This suggests that VR-SBE features extracted on the fly can instantaneously capture homogeneous characteristics of dysarthric speakers. The real-time factor (RTF) represents the overall delay incurred by waiting for data and model processing. By using only 10 ms of acoustic inputs, our VR-SBE adaptation achieves both stat. significant improvements over iVector or xVector adaptation and lower processing latency with a real-time factor speeding up by 33.6 times (Sys.10 vs. Sys.1-2).

\vspace{-0.5em}
\subsection{Experiments and Results on Elderly Speech}
\label{sec:exp_elderly}

\subsubsection{The DementiaBank Pitt Dataset}
\label{sec:exp_elderly_dbank_intro}


The English DementiaBank Pitt~\cite{becker1994natural} dataset contains 33h of cognitive impairment assessment interviews between 292 elderly participants and clinical investigators. The training set includes 688 speakers (244 elderly and 444 investigators), while the development and evaluation sets\renewcommand\footnoterule{\vspace{-8pt}}\footnote{The evaluation set contains the same 48 speakers' Cookie theft picture description recordings as the test set of ADReSS~\cite{luz2020alzheimer} while the development set contains recordings of these speakers in other tasks if available.} respectively contain 119 (43 elderly and 76 investigators) and 95 speakers (48 elderly and 47 investigators). There is no overlap speaker between the training and the development or evaluation sets. Silence stripping~\cite{ye2021development} leads to a 15.7h training set (29682 utt.), a 2.5h development set (5103 utt.) and a 0.6h evaluation set (928 utt.), while data augmentation~\cite{ye2021development} produces a 58.9h training set (112830 utt.).

\subsubsection{The JCCOCC MoCA Dataset}
\label{sec:exp_elderly_jm_intro}

The Cantonese JCCOCC MoCA dataset comprises cognitive impairment assessment interviews between 256 elderly participants and clinical investigators~\cite{xu2021speaker}. The training set contains 369 speakers (158 elderly and 211 investigators), while the development and evaluation sets each contain speech from 49 elderly other than those in the training set. Silence stripping leads to a 32.1h training set (95448 utt.), a 3.5h development set (13675 utt.) and a 3.4h evaluation set (13414 utt.). Further data augmentation~\cite{geng2021spectro} produces a 156.9h training set (389049 utt.).

\subsubsection{Experiment Setup for the DementiaBank Pitt Corpus}
\label{sec:exp_elderly_dbank_setup}

Following the Kaldi~\cite{povey2011kaldi} chain setup, the hybrid factorized TDNN systems contain 14 context-slicing layers with a 3-frame context. The setup of E2E graphemic Conformers follows that for UASpeech. $40$-dim FBK features are used as inputs. A word-level 4-gram LM with Kneser-Ney (KN) smoothing is trained~\cite{ye2021development}, with a 3.8k vocabulary covering all words in DementiaBank Pitt adopted during evaluation.

\subsubsection{Experiment Setup for the JCCOCC MoCA Corpus}

\label{sec:exp_elderly_jm_setup}

The setup of the hybrid TDNN and E2E character Conformer systems are the same as those for DementiaBank Pitt. 40-dim FBK features are adopted as inputs. A word-level 4-gram LM with KN smoothing is trained using the transcription of JCCOCC MoCA (610k words), with a 5.2k recognition vocabulary covering all words in JCCOCC MoCA adopted.

\subsubsection{Performance Analyses}
\label{sec:exp_elderly_analysis}

\begin{table}[htbp]
\vspace{-5pt}
    \caption{Performance comparison between baseline iVector, xVector, batch-mode LHUC or SBE~\cite{geng2022spectro} adaptation and the proposed VR-SBE on-the-fly speaker adaptation on the \textbf{DementiaBank Pitt} corpus. ``Dev'' and ``Eval'' stand for the development and evaluation sets. ``INV'' and ``PAR'' refer to clinical investigator and elderly participant.  $^\star$ denotes a stat. significant improvement ($\alpha=0.05$) obtained over both iVector (Sys.2,10,14) and xVector (Sys.3,11,15) adaptation.}
    \label{tab:recog-DBANK}
    \centering
    \renewcommand\arraystretch{1.0}
    \renewcommand\tabcolsep{2.0pt}
    \scalebox{0.63}{\begin{tabular}{c|c|c|c|c|c|cc|cc|c}
    \hline\hline  
        \multirow{3}{*}{Sys.} & 
        \multirow{3}{*}{\tabincell{c}{Model\\(\#Para.)}} & 
        \multirow{3}{*}{\tabincell{c}{Adapt.\\Feat.}} & 
        \multirow{3}{*}{\tabincell{c}{LHUC\\SAT}} &
        \multirow{3}{*}{f-LHUC} &
        \multirow{3}{*}{\tabincell{c}{On\\Fly}} &
        \multicolumn{5}{c}{WER\%} \\  
    \cline{7-11}
        & & & & & & \multicolumn{2}{c|}{Dev} & \multicolumn{2}{c|}{Eval} & \multirow{2}{*}{All (PAR)} \\
    \cline{7-10}
        & & & & & & INV & PAR & INV & PAR &  \\
    \hline\hline
        1 & \multirow{11}{*}{\tabincell{c}{Hybrid\\TDNN\\(18M)}} & \xmark & \multirow{6}{*}{\xmark} & \multirow{6}{*}{\xmark} & - & 19.91 & 47.93 & 19.76 & 36.66 & 33.80 (44.59) \\
        2 & & iVector & & & \cmark & 19.97 & 46.76 & 18.20 & 37.01 & 33.37 (43.87) \\
        3 & & xVector & & & \cmark & 19.75 & 47.58 & 19.31 & 35.57 & 33.40 (44.02) \\
        4 & & SBE~\cite{geng2022spectro} & & & \xmark & 18.61 & 43.84 & 17.98 & 33.82 & 31.12 (40.87)\\
        5 & & iVR & & & \cmark & 19.19 & 47.64 & 18.65 & 35.80 & 33.26 (44.13) \\
        6 & & VR-SBE & & & \cmark & 18.72$^\star$ & 44.67$^\star$ & 18.65 & 34.03$^\star$ & \textbf{31.55$^\star$ (41.52)$^\star$}  \\
    \cline{1-1}\cline{3-11}
        7 & & \xmark & \multirow{2}{*}{\cmark} & \multirow{2}{*}{\xmark} & \xmark & 19.26 & 45.49 & 18.42 & 35.44 & 32.33 (42.51) \\
        8 & & SBE~\cite{geng2022spectro} & & & \xmark & 17.41 & 40.94 & 17.98 & 31.89 & 29.16 (38.26) \\
    \cline{1-1}\cline{3-11}
        9 & & \xmark & \multirow{3}{*}{\xmark} & (FBK) & \cmark & 19.61 & 45.40 & 18.87 & 34.77 & 32.33 (42.25) \\
        10 & & iVector & & (+iVector) & \cmark & 18.75 & 47.07 & 17.98 & 36.11 & 32.85 (43.83) \\
        11 & & xVector & & (+xVector) & \cmark & 19.50 & 46.23 & 18.98 & 35.53 & 32.75 (43.06) \\
        12 & & VR-SBE & & (+VR-SBE) & \cmark & 17.87$^\star$ & 43.83$^\star$ & 16.87$^\star$ & 34.56$^\star$ & \textbf{30.91$^\star$ (41.09)$^\star$} \\
    \cline{1-1}\cline{3-11}
        6+12 & & \multicolumn{3}{c|}{-} & \cmark & 17.66$^\star$ & \textbf{43.48$^\star$} & 16.09$^\star$ & \textbf{33.68$^\star$} & \textbf{30.51$^\star$ (40.58)$^\star$} \\
    \hline\hline 
        13 & \multirow{6}{*}{\tabincell{c}{CONF.\\(52M)}} & \xmark & \multirow{6}{*}{\xmark} & \multirow{6}{*}{\xmark} & - & 20.97 & 48.71 & 19.42 & 36.93 & 34.57 (45.22) \\
        14 & & iVector & & & \cmark & 21.48 & 48.32 & 17.42 & 37.79 & 34.71 (45.20) \\
        15 & & xVector & & & \cmark & 21.77 & 49.38 & 18.09 & 37.83 & 35.27 (45.96) \\
        16 & & SBE~\cite{geng2022spectro} & & & \xmark & 20.44 & 47.70 & 17.31 & 36.11 & 33.76 (44.27) \\
        17 & & iVR & & & \cmark & 22.09 & 49.56 & 19.64  & 38.58 & 35.65 (46.31) \\
        18 & & VR-SBE & & & \cmark & 20.83$^\star$ & \textbf{47.39$^\star$} & 17.64 & \textbf{36.34$^\star$} & \textbf{33.84$^\star$ (44.12)$^\star$} \\
    \hline\hline     
    \end{tabular}}
\vspace{-5pt}
\end{table}

The performance comparison between our on-the-fly VR-SBE adaptation, iVector~\cite{saon2013speaker} or xVector~\cite{snyder2018x} adaptation and the batch-mode LHUC or SBE adaptation on DementiaBank Pitt is demonstrated in Table~\ref{tab:recog-DBANK}. Several trends can be observed: 

i) The proposed on-the-fly VR-SBE adapted systems (Sys.6,18) consistently outperform the iVector (Sys.2,14) and xVector adaptation (Sys.3,15) on hybrid TDNN and E2E Conformer systems. Statistically significant overall WER reductions by up to \textbf{1.85\% abs. (5.54\% rel.)} and 1.43\% abs. (4.05\% relative) are respectively obtained on TDNN (Sys.6 vs. Sys.3) and Conformer (Sys.18 vs. Sys.15) models. 

ii) On-the-fly VR-SBE adaptation (Sys.6,18) consistently outperforms the comparable variance-regularized iVector (iVR) adaptation (Sys.5,17). 

iii) On-the-fly VR-SBE adaptation also outperforms batch-mode LHUC adaptation with a WER reduction of 0.78\% abs. (2.41\% rel.) (Sys.6 vs. Sys.7). 

iv) On-the-fly VR-SBE adaptation (Sys.6,18) produces performance comparable to batch-mode SBE adaptation~\cite{geng2022spectro} using time-consuming speaker embedding averaging (Sys.4,16). 

v) On-the-fly VR-SBE plus FBK+VR-SBE driven f-LHUC adaptation (Sys.12) not only outperforms the comparable baselines replacing VR-SBE with iVector or xVector (Sys.10-11), but also gives further WER reductions over VR-SBE adaptation alone by 0.64\% abs. (2.03\% rel.) (Sys.12 vs. Sys.6). 

vi) Combining VR-SBE plus FBK+VR-SBE feature driven f-LHUC adaptation (Sys.12) and VR-SBE adaptation alone (Sys.6) via frame-level joint decoding (Sys.6+12)\renewcommand\footnoterule{\vspace{-8pt}}\footnote{System weights empirically set as 11:9.} gives the best performance, with statistically significant overall WER reductions by up to \textbf{2.89\% abs. (8.65\% rel.)} over iVector and xVector adaptation (Sys.6+12 vs. Sys.3).

\begin{table}[htbp]
\vspace{-5pt}
    \caption{Performance comparison between baseline iVector, xVector, batch-mode LHUC or SBE~\cite{geng2022spectro} adaptation and the proposed VR-SBE on-the-fly speaker adaptation on the \textbf{JCCOCC MoCA} development (Dev) and evaluation (Eval) sets containing elderly speakers only. $^\star$ denotes a stat. significant improvement ($\alpha=0.05$) obtained over both iVector (Sys.2,10,14) and xVector (Sys.3,11,15) adaptation.}
    \label{tab:recog-JM}
    \centering
    \renewcommand\arraystretch{1.0}
    \renewcommand\tabcolsep{2.0pt}
    \scalebox{0.78}{\begin{tabular}{c|c|c|c|c|c|c|c|c}
    \hline\hline  
        \multirow{2}{*}{Sys.} & 
        \multirow{2}{*}{\tabincell{c}{Model\\(\#Para.)}} & 
        \multirow{2}{*}{\tabincell{c}{Adapt.\\Feat.}} & 
        \multirow{2}{*}{\tabincell{c}{LHUC}} &
        \multirow{2}{*}{f-LHUC} &
        \multirow{2}{*}{\tabincell{c}{On\\Fly}} &
        \multicolumn{3}{c}{CER\%} \\  
    \cline{7-9}
        & & & & & & Dev & Eval & All \\ 
    \hline\hline
        1 & \multirow{11}{*}{\tabincell{c}{Hybrid\\TDNN\\(18M)}} & \xmark & \multirow{6}{*}{\xmark} & \multirow{6}{*}{\xmark} & - & 26.87 & 23.71 & 25.28 \\
        2 & & iVector & & & \cmark & 25.46 & 22.80 & 24.12 \\
        3 & & xVector & & & \cmark & 26.72 & 23.80 & 25.25 \\
        4 & & SBE~\cite{geng2022spectro} & & & \xmark & 24.43 & 21.68 & 23.05 \\
        5 & & iVR & & & \cmark & 25.22 & 22.74 & 23.97  \\
        6 & & VR-SBE & & & \cmark & 24.86$^{\star}$ & 22.18$^{\star}$ & \textbf{23.51$^{\star}$}  \\
    \cline{1-1}\cline{3-9}
        7 & & \xmark & \multirow{2}{*}{\cmark} & \multirow{2}{*}{\xmark} & \xmark & 25.77 & 22.94 & 24.35 \\
        8 & & SBE~\cite{geng2022spectro} & & & \xmark & 23.59 & 21.42 & 22.50  \\
    \cline{1-1}\cline{3-9}
        9 & & \xmark & \multirow{3}{*}{\xmark} & (FBK) & \cmark & 24.93 & 21.82 & 23.37  \\
        10 & & iVector & & (+iVector) & \cmark & 25.03 & 22.58 & 23.80 \\
        11 & & xVector & & (+xVector) & \cmark & 25.42 & 22.28 & 23.84 \\
        12 & & VR-SBE & & (+VR-SBE) & \cmark  & 23.59$^\star$ & 21.12$^\star$ & \textbf{22.35$^\star$}  \\
    \cline{1-1}\cline{3-9}
        6+12 & & \multicolumn{3}{c|}{-} & \cmark & \textbf{23.55$^\star$} & \textbf{20.70$^\star$} & \textbf{22.11$^\star$} \\
    \hline\hline 
        13 & \multirow{6}{*}{\tabincell{c}{CONF.\\(53M)}} & \xmark & \multirow{6}{*}{\xmark} & \multirow{6}{*}{\xmark} & - &   33.08 & 31.24 & 32.15 \\
        14 & & iVector & & & \cmark & 33.76 & 31.83 & 32.79 \\
        15 & & xVector & & & \cmark & 33.43 & 32.10 & 32.76 \\
        16 & & SBE~\cite{geng2022spectro} & & & \xmark & 32.08 & 30.75 & 31.41 \\
        17 & & iVR & & & \cmark & 34.79 & 32.48 & 33.63 \\
        18 & & VR-SBE & & & \cmark & \textbf{32.42$^\star$} & \textbf{31.01$^\star$} & \textbf{31.71$^\star$} \\
    \hline\hline     
    \end{tabular}}
\vspace{-5pt}
\end{table}

Experiments conducted on JCCOCC MoCA~\cite{xu2021speaker} (Table~\ref{tab:recog-JM}) demonstrate similar trends. On-the-fly VR-SBE adaptation (Sys.6,18) outperforms iVector (Sys.2,14) or xVector (Sys.3,15) adaptation by up to \textbf{1.74\% abs. (6.89\% rel.)} CER reduction (Sys.6 vs. Sys.3), and batch-mode LHUC adaptation by \textbf{0.84\% abs. (3.45\% rel.)} CER reduction (Sys.6 vs. Sys.7). Combining on-the-fly VR-SBE and FBK+VR-SBE driven f-LHUC adaptation leads to CER reductions of up to \textbf{3.14\% abs. (12.44\% rel.)} over iVector and xVector adaptation (Sys.6+12 vs. Sys.3) and \textbf{2.24\% abs.(9.20\% rel.)} over batch-mode LHUC adaptation (Sys.6+12 vs. Sys.7).

\subsubsection{Processing Latency Analyses}
\label{sec:exp_elderly_latency}

\begin{table}[htbp]
\vspace{0pt}
    \caption{Ablation study of the analysis sliding window (Slid. Wind.) size for on-the-fly VR-SBE feature extraction on \textbf{DimentiaBank Pitt}. ``RTF'' denotes real-time factor. $^\dag$, $^\ddag$ and $^\star$ denote a stat. significant improvement ($\alpha=0.05$) is obtained over iVector (Sys.1), xVector (Sys.2), or both.}
    \label{tab:VR-SBE-sliding-window-DBANK}
    \centering
    \renewcommand\arraystretch{1.0}
    \renewcommand\tabcolsep{2.0pt}
    \scalebox{0.67}{\begin{tabular}{c|c|c|c|c|cc|cc|c}
    \hline\hline  
        \multirow{3}{*}{Sys.} & 
        \multirow{3}{*}{\tabincell{c}{Model\\(\#Para.)}} & 
        \multirow{3}{*}{\tabincell{c}{Adapt.\\Feat.}} & 
        \multirow{3}{*}{\tabincell{c}{Slid.\\Wind.}} &
        \multirow{3}{*}{RTF} &
        \multicolumn{5}{c}{WER\%} \\
    \cline{6-10}
         & & & & & \multicolumn{2}{c|}{Dev} & \multicolumn{2}{c|}{Eval} & \multirow{2}{*}{All (PAR)} \\
    \cline{6-9}
         & & & & & INV & PAR & INV & PAR & \\
    \hline\hline
        1 & \multirow{10}{*}{\tabincell{c}{Hybrid\\TDNN\\(18M)}} & iVector & 100ms & 0.08 & 19.97 & 46.76 & 18.20 & 37.01 & 33.37 (43.87) \\
        2 & & xVector & utt. & 1.02 & 19.75 & 47.58 & 19.31 & 35.57 & 33.40 (44.02) \\
    \cline{1-1} \cline{3-10}
        3 & & \multirow{8}{*}{VR-SBE} & utt. & 1.03 & 18.72$^\star$ & 44.67$^\star$ & 18.65 & 34.03$^\star$ & 31.55$^\star$ (41.52)$^\star$ \\
        4 & & & 250ms & 0.16 & 18.88$^\star$ & 44.91$^\star$ & 18.98 & 34.37$^\star$ & 31.78$^\star$ (41.79)$^\star$  \\
        5 & & & 150ms & 0.11 & 19.21$^\dag$ & 44.83$^\star$ & 17.87 & 33.53$^\star$ & 31.71$^\star$ (41.48)$^\star$ \\
        6 & & & 100ms & 0.08 & 19.08$^\star$ & 45.03$^\star$ & 19.42 & 34.41$^\star$ & 31.93$^\star$ (41.89)$^\star$ \\
        7 & & & 50ms & 0.06 & 19.38 & 44.87$^\star$ & 18.31$^\ddag$  & 34.52$^\star$ & 31.97$^\star$ (41.81)$^\star$ \\
        8 & & & 30ms & 0.05 & 18.93$^\star$ & 44.57$^\star$ & 18.53 & 34.52$^\star$ & 31.68$^\star$ (41.59)$^\star$ \\
        9 & & & 20ms & 0.04 & 18.91$^\star$ & 45.45$^\star$ & 18.20$^\ddag$ & 34.10$^\star$ & 31.94$^\star$ (42.09)$^\star$ \\
        10 & & & 10ms & 0.04 & 19.33 & \textbf{44.93$^\star$} & 19.09 & \textbf{34.35$^\star$} & \textbf{31.97$^\star$ (41.80)$^\star$} \\
    \hline\hline
    \end{tabular}}
\vspace{-0pt}
\end{table}


The results in Table~\ref{tab:VR-SBE-sliding-window-DBANK} show trends similar to those previously found in Table~\ref{tab:VR-SBE-sliding-window-UASpeech} on UASpeech, and further confirm that the performance of on-the-fly VR-SBE adaptation is largely insensitive to changes in the size of input feature analysis sliding window.

\vspace{-1em}
\subsection{Further Analyses on Feature Homogeneity}

\begin{figure}[htbp]
  \centering
  \includegraphics[scale=0.18]{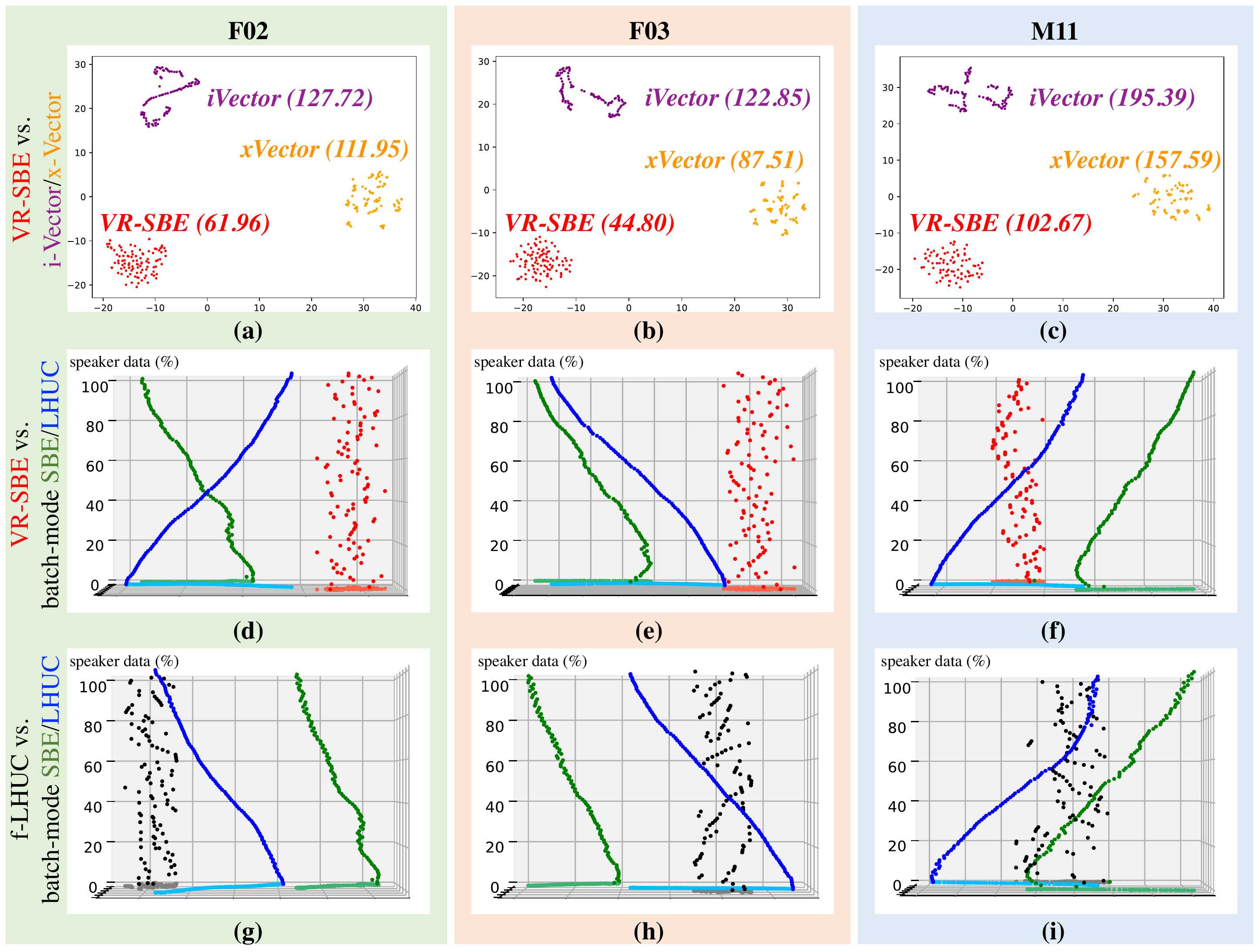}
  \caption{T-SNE visualization illustrating speaker feature homogeneity measured by covariance determinants after applying t-SNE projection: \textbf{(a)-(c)} on-the-fly VR-SBE features vs. iVectors and xVectors \textbf{(d)-(f)} VR-SBE vs. batch-mode SBE features and LHUC transforms and \textbf{(g)-(i)} VR-SBE feature driven f-LHUC vs. batch-mode SBE features and LHUC transforms obtained from speaker F02, F03 and M11 of \textbf{UASpeech}.}
  \vspace{-\baselineskip} 
  \label{fig:t-SNE}
\end{figure} 

We further analyze the homogeneity of the proposed on-the-fly adaptation features. T-distributed stochastic neighbor embedding (t-SNE)~\cite{van2008visualizing} visualization is conducted on the proposed VR-SBE features against baseline iVectors and x-Vectors in 2-D plots (Fig.~\ref{fig:t-SNE}(a)-(c)), and also on VR-SBE or FBK+VR-SBE feature driven f-LHUC transforms against batch-mode spectral basis embedding (SBE) features and LHUC transforms in 3-D plots (Fig.~\ref{fig:t-SNE}(d)-(i)). Three UASpeech\renewcommand\footnoterule{\vspace{-8pt}}\footnote{The amounts of speaker-level data of test speakers of UASpeech (34 min, including all channels) are larger than those of TORGO (7.8 min), DementiaBank Pitt (0.4 min) and JCCOCC MoCA (4.2 min). Given this, t-SNE visualizations are conducted using the UASpeech data.} dysarthric speakers of mixed genders are selected from the very low (F03), low (F02) and mid (M11) speech intelligibility groups. For each speaker, 101 distinct speaker-level adaptation data quantity operating points are used for computing the above speaker features or LHUC transforms. These points correspond to using only one utterance or 1\%, 2\%,...,99\%, and up to 100\% of the speaker-level data.  Fig.~\ref{fig:t-SNE} shows that the proposed on-the-fly VR-SBE features and their associated f-LHUC transforms consistently exhibit stronger speaker homogeneity measured in covariance determinants after applying t-SNE projection than baseline iVectors, x-Vectors and batch-mode SBE features and LHUC transforms.

\begin{figure}
   \centering
    \subfloat[\textbf{UASpeech}]{\includegraphics[width=0.36\textwidth]{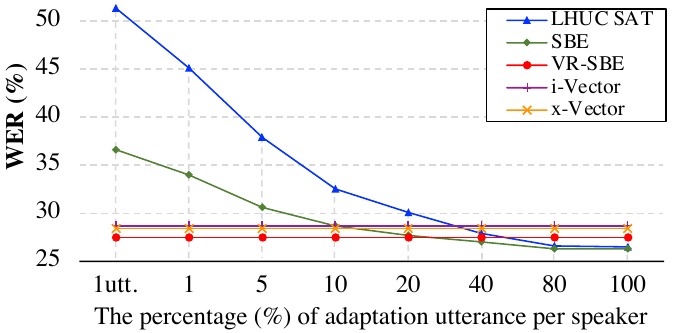}\label{fig:limited-amount-UA}}  
    \vspace{0em}
    \subfloat[\textbf{DementiaBank Pitt}]{\includegraphics[width=0.36\textwidth]{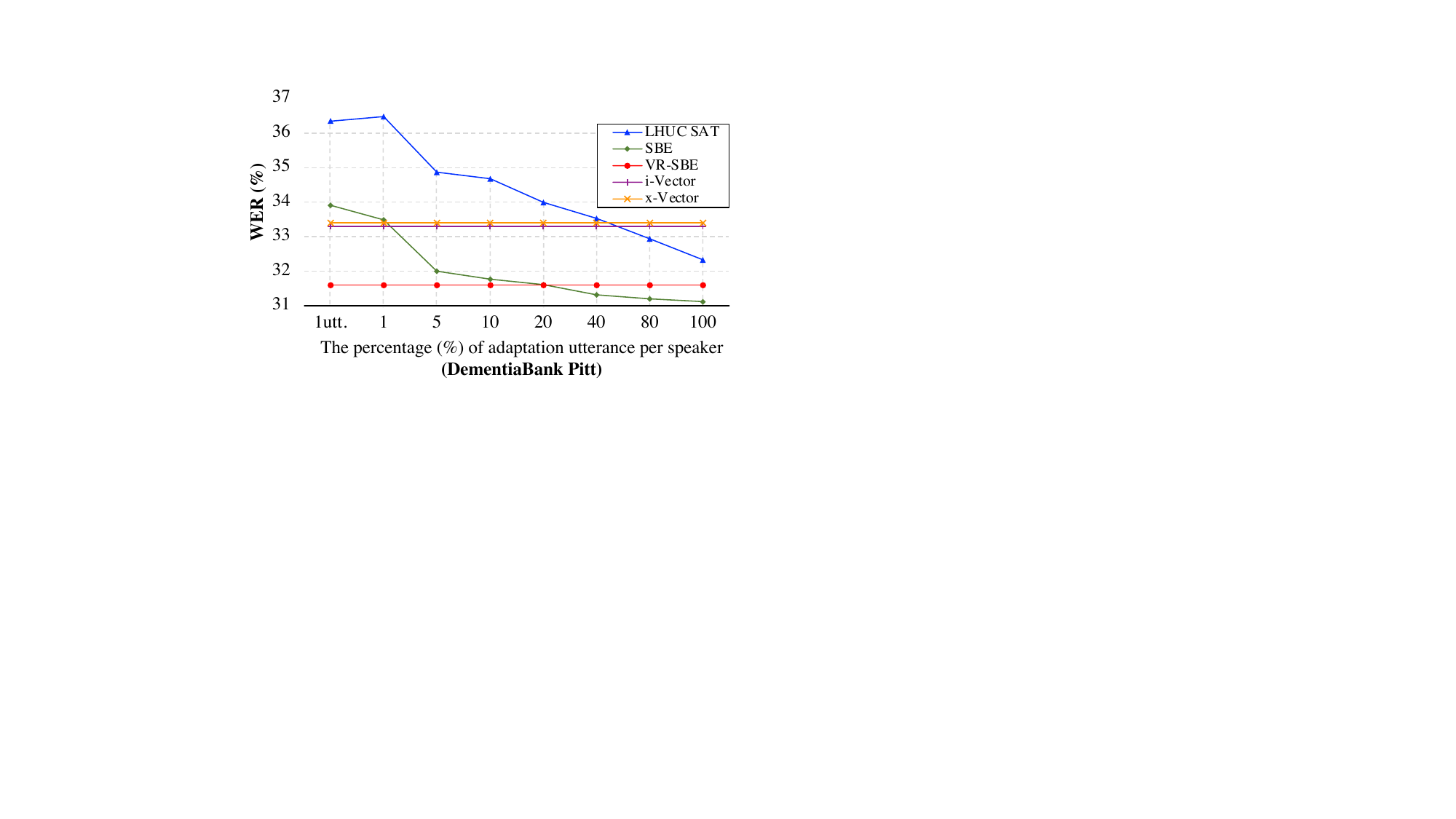}\label{fig:limited-amount-DBANK}}
  \caption{Performance comparison of hybrid DNN/TDNN systems between batch-mode LHUC/SBE adaptation, on-the-fly iVector/x-Vector adaptation, and the proposed on-the-fly VR-SBE adaptation with various percentages of speaker-level adaptation data on \textbf{(a)} UASpeech and \textbf{(b)} DementiaBank Pitt.}
  \vspace{-\baselineskip} 
  \label{fig:limited-amount}
\end{figure}

We further compare the performance of the proposed on-the-fly VR-SBE adaptation with batch-mode LHUC and SBE adaptation when using limited amounts of speaker-level adaptation data. As depicted in Fig.~\ref{fig:limited-amount}(a) and Fig.~\ref{fig:limited-amount}(b), VR-SBE adaptation is more robust to changes in the amount of speaker-level data than batch-mode LHUC and SBE adaptation. Specifically, VR-SBE adaptation outperforms SBE adaptation when using less than 20\% of the speaker data and surpasses LHUC adaptation when using less than 40\% of speaker data. 

\section{Discussion and Conclusion}
\label{sec:conclusion}


This paper presents two novel methods to learn homogeneous dysarthric and elderly speaker features for on-the-fly test-time adaptation of TDNN and Conformer models,
including: 1) speaker-level variance-regularized spectral basis
embedding (VR-SBE) features that ensure speaker-level feature consistency 
via specially designed regularization; and feature-based LHUC
(f-LHUC) transforms driven by VR-SBE. Experiments
conducted on four dysarthric and elderly speech corpora across English and Cantonese suggest our proposed approaches
achieve statistically significant WER or CER reductions of up
to 5.32\% absolute (18.57\% relative) over baseline iVector/xVector adaptation, and 2.24\% absolute (9.20\% relative) over batch-mode offline LHUC adaptation. 
Further processing latency analyses and t-SNE visualization show that our VR-SBE features and f-LHUC transforms are robust to speaker-level data quantity in test-time adaptation, while exhibiting stronger speaker-level homogeneity than iVectors, xVectors
and batch-mode LHUC transforms. Future research will focus
on rapid adaptation of pre-trained ASR systems.

\ifCLASSOPTIONcaptionsoff
  \newpage
\fi

\bibliographystyle{IEEEtran}
\bibliography{main}

\begin{thebibliography}{100}
\providecommand{\url}[1]{#1}
\csname url@samestyle\endcsname
\providecommand{\newblock}{\relax}
\providecommand{\bibinfo}[2]{#2}
\providecommand{\BIBentrySTDinterwordspacing}{\spaceskip=0pt\relax}
\providecommand{\BIBentryALTinterwordstretchfactor}{4}
\providecommand{\BIBentryALTinterwordspacing}{\spaceskip=\fontdimen2\font plus
\BIBentryALTinterwordstretchfactor\fontdimen3\font minus \fontdimen4\font\relax}
\providecommand{\BIBforeignlanguage}[2]{{%
\expandafter\ifx\csname l@#1\endcsname\relax
\typeout{** WARNING: IEEEtran.bst: No hyphenation pattern has been}%
\typeout{** loaded for the language `#1'. Using the pattern for}%
\typeout{** the default language instead.}%
\else
\language=\csname l@#1\endcsname
\fi
#2}}
\providecommand{\BIBdecl}{\relax}
\BIBdecl

\bibitem{christensen2012comparative}
H.~Christensen \emph{et~al.}, ``{A comparative study of adaptive, automatic recognition of disordered speech},'' in \emph{INTERSPEECH}, 2012.

\bibitem{sehgal2015model}
S.~Sehgal \emph{et~al.}, ``{Model adaptation and adaptive training for the recognition of dysarthric speech},'' in \emph{SLPAT}, 2015.

\bibitem{hu2019cuhk}
S.~Hu \emph{et~al.}, ``{The CUHK Dysarthric Speech Recognition Systems for English and Cantonese},'' in \emph{INTERSPEECH}, 2019.

\bibitem{shahamiri2021speech}
S.~R. Shahamiri, ``{Speech vision: An end-to-end deep learning-based dysarthric automatic speech recognition system},'' \emph{IEEE T NEUR SYS REH}, 2021.

\bibitem{baskar2022speaker}
M.~K. Baskar \emph{et~al.}, ``{Speaker adaptation for Wav2vec2 based dysarthric ASR},'' in \emph{INTERPSEECH}, 2022.

\bibitem{hernandez22_interspeech}
A.~Hernandez \emph{et~al.}, ``{Cross-lingual Self-Supervised Speech Representations for Improved Dysarthric Speech Recognition},'' in \emph{INTERSPEECH}, 2022.

\bibitem{violeta2022investigating}
L.~P. Violeta \emph{et~al.}, ``{Investigating Self-supervised Pretraining Frameworks for Pathological Speech Recognition},'' in \emph{INTERSPEECH}, 2022.

\bibitem{hu2022exploiting}
S.~Hu \emph{et~al.}, ``{Exploiting Cross-domain And Cross-Lingual Ultrasound Tongue Imaging Features For Elderly And Dysarthric Speech Recognition},'' in \emph{INTERSPEECH}, 2023.

\bibitem{almadhor2023e2e}
A.~Almadhor \emph{et~al.}, ``{E2E-DASR: End-to-end deep learning-based dysarthric automatic speech recognition},'' \emph{EXPERT SYST APPL}, 2023.

\bibitem{kim2023unsupervised}
J.-W. Kim \emph{et~al.}, ``{Unsupervised Representation Learning with Task-Agnostic Feature Masking for Robust End-to-End Speech Recognition},'' \emph{Mathematics}, 2023.

\bibitem{jin2022personalized}
Z.~Jin \emph{et~al.}, ``{Personalized Adversarial Data Augmentation for Dysarthric and Elderly Speech Recognition},'' \emph{IEEE T AUDIO SPEECH}, 2023.

\bibitem{lanier2010speech}
W.~Lanier, \emph{{Speech disorders}}.\hskip 1em plus 0.5em minus 0.4em\relax Greenhaven Publishing LLC, 2010.

\bibitem{fraser2016linguistic}
K.~C. Fraser \emph{et~al.}, ``{Linguistic features identify Alzheimer’s disease in narrative speech},'' \emph{J. Alzheimer's Dis.}, 2016.

\bibitem{alzheimer20192019}
A.~Association, ``{2019 Alzheimer's disease facts and figures},'' \emph{Alzheimer's \& dementia}, 2019.

\bibitem{young2010difficulties}
V.~Young \emph{et~al.}, ``{Difficulties in automatic speech recognition of dysarthric speakers and implications for speech-based applications used by the elderly: A literature review},'' \emph{ASSIST TECHNOL}, 2010.

\bibitem{vipperla2010ageing}
R.~Vipperla \emph{et~al.}, ``{Ageing voices: The effect of changes in voice parameters on ASR performance},'' \emph{EURASIP J AUDIO SPEE}, 2010.

\bibitem{christensen2013combining}
H.~Christensen \emph{et~al.}, ``{Combining in-domain and out-of-domain speech data for automatic recognition of disordered speech},'' in \emph{INTERSPEECH}, 2013.

\bibitem{zhou2016speech}
L.~Zhou \emph{et~al.}, ``{Speech Recognition in Alzheimer's Disease and in its Assessment},'' in \emph{INTERSPEECH}, 2016.

\bibitem{vachhani2017deep}
B.~Vachhani \emph{et~al.}, ``{Deep Autoencoder Based Speech Features for Improved Dysarthric Speech Recognition},'' in \emph{INTERSPEECH}, 2017.

\bibitem{kim2018dysarthric}
M.~J. Kim \emph{et~al.}, ``{Dysarthric Speech Recognition Using Convolutional LSTM Neural Network},'' in \emph{INTERSPEECH}, 2018.

\bibitem{joy2018improving}
N.~M. Joy \emph{et~al.}, ``{Improving acoustic models in TORGO dysarthric speech database},'' \emph{IEEE T NEUR SYS REH}, 2018.

\bibitem{liu2019exploiting}
S.~Liu \emph{et~al.}, ``{Exploiting Visual Features Using Bayesian Gated Neural Networks for Disordered Speech Recognition},'' in \emph{INTERSPEECH}, 2019.

\bibitem{Shor2019}
J.~Shor \emph{et~al.}, ``{Personalizing ASR for Dysarthric and Accented Speech with Limited Data},'' in \emph{INTERSPEECH}, 2019.

\bibitem{geng2020investigation}
M.~Geng \emph{et~al.}, ``{Investigation of Data Augmentation Techniques for Disordered Speech Recognition},'' in \emph{INTERSPEECH}, 2020.

\bibitem{lin2020staged}
Y.~Lin \emph{et~al.}, ``{Staged Knowledge Distillation for End-to-End Dysarthric Speech Recognition and Speech Attribute Transcription},'' in \emph{INTERSPEECH}, 2020.

\bibitem{kodrasi2020spectro}
I.~Kodrasi \emph{et~al.}, ``{Spectro-Temporal Sparsity Characterization for Dysarthric Speech Detection},'' \emph{IEEE T AUDIO SPEECH}, 2020.

\bibitem{xiong2020source}
F.~Xiong \emph{et~al.}, ``{Source Domain Data Selection for Improved Transfer Learning Targeting Dysarthric Speech Recognition},'' in \emph{ICASSP}, 2020.

\bibitem{takashima2020two}
R.~Takashima \emph{et~al.}, ``{Two-step acoustic model adaptation for dysarthric speech recognition},'' in \emph{ICASSP}, 2020.

\bibitem{liu2021recent}
S.~Liu \emph{et~al.}, ``{Recent Progress in the CUHK Dysarthric Speech Recognition System},'' \emph{IEEE T AUDIO SPEECH}, 2021.

\bibitem{jin2021adversarial}
Z.~Jin \emph{et~al.}, ``{Adversarial Data Augmentation for Disordered Speech Recognition},'' in \emph{INTERSPEECH}, 2021.

\bibitem{hu2021bayesian}
S.~Hu \emph{et~al.}, ``{Bayesian Learning of LF-MMI Trained Time Delay Neural Networks for Speech Recognition},'' \emph{IEEE T AUDIO SPEECH}, 2021.

\bibitem{wang2021improved}
D.~Wang \emph{et~al.}, ``{Improved end-to-end dysarthric speech recognition via meta-learning based model re-initialization},'' in \emph{ISCSLP}, 2021.

\bibitem{green2021automatic}
J.~R. Green \emph{et~al.}, ``{Automatic Speech Recognition of Disordered Speech: Personalized models outperforming human listeners on short phrases},'' in \emph{INTERSPEECH}, 2021.

\bibitem{ye2021development}
Z.~Ye \emph{et~al.}, ``{Development of the CUHK Elderly Speech Recognition System for Neurocognitive Disorder Detection Using the Dementiabank Corpus},'' in \emph{ICASSP}, 2021.

\bibitem{pan2021using}
Y.~Pan \emph{et~al.}, ``{Using the Outputs of Different Automatic Speech Recognition Paradigms for Acoustic-and BERT-Based Alzheimer’s Dementia Detection Through Spontaneous Speech},'' in \emph{INTERSPEECH}, 2021.

\bibitem{geng2021spectro}
M.~Geng \emph{et~al.}, ``{Spectro-Temporal Deep Features for Disordered Speech Assessment and Recognition},'' in \emph{INTERSPEECH}, 2021.

\bibitem{harvill2021synthesis}
J.~Harvill \emph{et~al.}, ``{Synthesis of new words for improved dysarthric speech recognition on an expanded vocabulary},'' in \emph{ICASSP}, 2021.

\bibitem{geng2022spectro}
M.~Geng \emph{et~al.}, ``{Speaker Adaptation Using Spectro-Temporal Deep Features for Dysarthric and Elderly Speech Recognition},'' \emph{IEEE T AUDIO SPEECH}, 2022.

\bibitem{hu2023exploring}
S.~Hu \emph{et~al.}, ``{Exploring Self-supervised Pre-trained ASR Models For Dysarthric and Elderly Speech Recognition},'' in \emph{ICASSP}, 2023.

\bibitem{geng2023fly}
M.~Geng \emph{et~al.}, ``{On-the-Fly Feature Based Rapid Speaker Adaptation for Dysarthric and Elderly Speech Recognition},'' in \emph{INTERSPEECH}, 2023.

\bibitem{jin2022adversarial}
Z.~Jin \emph{et~al.}, ``{Adversarial Data Augmentation Using VAE-GAN for Disordered Speech Recognition},'' in \emph{ICASSP}, 2023.

\bibitem{qi2023parameter}
J.~Qi \emph{et~al.}, ``{Parameter-efficient Dysarthric Speech Recognition Using Adapter Fusion and Householder Transformation},'' in \emph{INTRESPEECH}, 2023.

\bibitem{geng2023use}
M.~Geng \emph{et~al.}, ``{Use of Speech Impairment Severity for Dysarthric Speech Recognition},'' in \emph{INTERSPEECH}, 2023.

\bibitem{wang2024enhancing}
H.~Wang \emph{et~al.}, ``{Enhancing Pre-trained ASR System Fine-tuning for Dysarthric Speech Recognition using Adversarial Data Augmentation},'' in \emph{ICASSP}, 2024.

\bibitem{jerntorp1992stroke}
P.~Jerntorp \emph{et~al.}, ``{Stroke registry in Malm{\"o}, Sweden},'' \emph{STROKE}, 1992.

\bibitem{whitehill2000speech}
T.~L. Whitehill \emph{et~al.}, ``{Speech errors in Cantonese speaking adults with cerebral palsy},'' \emph{CLIN LINGUIST PHONET}, 2000.

\bibitem{caligiuri1989influence}
M.~P. Caligiuri, ``{The influence of speaking rate on articulatory hypokinesia in Parkinsonian dysarthria},'' \emph{BRAIN LANG}, 1989.

\bibitem{peppe2009prosody}
S.~J. Pepp{\'e}, ``{Why is prosody in speech-language pathology so difficult?}'' \emph{INT J SPEECH-LANG PA}, 2009.

\bibitem{abdel2013fast}
O.~Abdel-Hamid \emph{et~al.}, ``{Fast speaker adaptation of hybrid NN/HMM model for speech recognition based on discriminative learning of speaker code},'' in \emph{ICASSP}, 2013.

\bibitem{saon2013speaker}
G.~Saon \emph{et~al.}, ``{Speaker adaptation of neural network acoustic models using i-vectors},'' in \emph{ASRU}, 2013.

\bibitem{senior2014improving}
A.~Senior \emph{et~al.}, ``{Improving DNN speaker independence with i-vector inputs},'' in \emph{ICASSP}, 2014.

\bibitem{gales1998maximum}
M.~J. Gales, ``{Maximum likelihood linear transformations for HMM-based speech recognition},'' \emph{COMPUT SPEECH LANG}, 1998.

\bibitem{uebel1999investigation}
L.~F. Uebel \emph{et~al.}, ``{An investigation into vocal tract length normalisation},'' in \emph{EUROSPEECH}, 1999.

\bibitem{seide2011feature}
F.~Seide \emph{et~al.}, ``{Feature engineering in context-dependent deep neural networks for conversational speech transcription},'' in \emph{ASRU}, 2011.

\bibitem{gemello2007linear}
R.~Gemello \emph{et~al.}, ``{Linear hidden transformations for adaptation of hybrid ANN/HMM models},'' \emph{SPEECH COMMUN}, 2007.

\bibitem{li2010comparison}
B.~Li \emph{et~al.}, ``{Comparison of discriminative input and output transformations for speaker adaptation in the hybrid NN/HMM systems},'' in \emph{INTERSPEECH}, 2010.

\bibitem{swietojanski2016learning}
P.~Swietojanski \emph{et~al.}, ``{Learning hidden unit contributions for unsupervised acoustic model adaptation},'' \emph{IEEE T AUDIO SPEECH}, 2016.

\bibitem{zhang2016dnn}
C.~Zhang \emph{et~al.}, ``{DNN speaker adaptation using parameterised sigmoid and ReLU hidden activation functions},'' in \emph{ICASSP}, 2016.

\bibitem{sharma2010state}
H.~V. Sharma \emph{et~al.}, ``{State-transition interpolation and MAP adaptation for HMM-based dysarthric speech recognition},'' in \emph{SLPAT}, 2010.

\bibitem{baba2001elderly}
A.~Baba \emph{et~al.}, ``{Elderly acoustic model for large vocabulary continuous speech recognition},'' in \emph{EUROSPEECH}, 2001.

\bibitem{mengistu2011adapting}
K.~T. Mengistu \emph{et~al.}, ``{Adapting acoustic and lexical models to dysarthric speech},'' in \emph{ICASSP}, 2011.

\bibitem{kim2013dysarthric}
M.~J. Kim \emph{et~al.}, ``{Dysarthric speech recognition using dysarthria-severity-dependent and speaker-adaptive models},'' in \emph{INTERSPEECH}, 2013.

\bibitem{bhat2016recognition}
C.~Bhat \emph{et~al.}, ``{Recognition of Dysarthric Speech Using Voice Parameters for Speaker Adaptation and Multi-Taper Spectral Estimation},'' in \emph{INTERSPEECH}, 2016.

\bibitem{kim2017regularized}
M.~Kim \emph{et~al.}, ``{Regularized speaker adaptation of KL-HMM for dysarthric speech recognition},'' \emph{IEEE T NEUR SYS REH}, 2017.

\bibitem{tobin2022personalized}
J.~Tobin \emph{et~al.}, ``{Personalized automatic speech recognition trained on small disordered speech datasets},'' in \emph{ICASSP}, 2022.

\bibitem{espana2016automatic}
C.~Espana-Bonet \emph{et~al.}, ``{Automatic speech recognition with deep neural networks for impaired speech},'' in \emph{IberSPEECH}, 2016.

\bibitem{yue2022acoustic}
Z.~Yue \emph{et~al.}, ``{Acoustic Modelling From Raw Source and Filter Components for Dysarthric Speech Recognition},'' \emph{IEEE T AUDIO SPEECH}, 2022.

\bibitem{deng2021bayesian}
J.~Deng \emph{et~al.}, ``{Bayesian Parametric and Architectural Domain Adaptation of LF-MMI Trained TDNNs for Elderly and Dysarthric Speech Recognition},'' in \emph{INTERSPEECH}, 2021.

\bibitem{xie2019fast}
X.~Xie \emph{et~al.}, ``{Fast DNN Acoustic Model Speaker Adaptation by Learning Hidden Unit Contribution Features},'' in \emph{INTERSPEECH}, 2019.

\bibitem{kim2008dysarthric}
H.~Kim \emph{et~al.}, ``{Dysarthric speech database for universal access research},'' in \emph{INTERSPEECH}, 2008.

\bibitem{rudzicz2012torgo}
F.~Rudzicz \emph{et~al.}, ``{The TORGO database of acoustic and articulatory speech from speakers with dysarthria},'' \emph{LANG RESOUR EVAL}, 2012.

\bibitem{becker1994natural}
J.~T. Becker \emph{et~al.}, ``{The natural history of Alzheimer's disease: description of study cohort and accuracy of diagnosis},'' \emph{ARCH NEUROL-CHICAGO}, 1994.

\bibitem{xu2021speaker}
S.~S. Xu \emph{et~al.}, ``{Speaker Turn Aware Similarity Scoring for Diarization of Speech-Based Cognitive Assessments},'' in \emph{APSIPA ASC}, 2021.

\bibitem{snyder2018x}
D.~Snyder \emph{et~al.}, ``{X-vectors: Robust dnn embeddings for speaker recognition},'' in \emph{ICASSP}, 2018.

\bibitem{pallet1990tools}
D.~S. Pallet \emph{et~al.}, ``{Tools for the analysis of benchmark speech recognition tests},'' in \emph{ICASSP}, 1990.

\bibitem{van2008visualizing}
L.~Van~der Maaten \emph{et~al.}, ``{Visualizing data using t-SNE},'' \emph{J MACH LEARN RES}, 2008.

\bibitem{tuske2021limit}
Z.~T{\"u}ske \emph{et~al.}, ``{On the limit of English conversational speech recognition},'' in \emph{INTERSPEECH}, 2021.

\bibitem{zeineldeen2022improving}
M.~Zeineldeen \emph{et~al.}, ``{Improving the training recipe for a robust conformer-based hybrid model},'' in \emph{INTERSPEECH}, 2022.

\bibitem{eide1996parametric}
E.~Eide \emph{et~al.}, ``{A parametric approach to vocal tract length normalization},'' in \emph{ICASSP}, 1996.

\bibitem{lee1996speaker}
L.~Lee \emph{et~al.}, ``{Speaker normalization using efficient frequency warping procedures},'' in \emph{ICASSP}, 1996.

\bibitem{neto1995speaker}
J.~Neto \emph{et~al.}, ``{Speaker-adaptation for hybrid HMM-ANN continuous speech recognition system},'' in \emph{EUROSPEECH}, 1995.

\bibitem{xie2021bayesian}
X.~Xie \emph{et~al.}, ``{Bayesian learning for deep neural network adaptation},'' \emph{IEEE T AUDIO SPEECH}, 2021.

\bibitem{zhao2016low}
Y.~Zhao \emph{et~al.}, ``{Low-rank plus diagonal adaptation for deep neural networks},'' in \emph{ICASSP}, 2016.

\bibitem{ochiai2018speaker}
T.~Ochiai \emph{et~al.}, ``{Speaker adaptation for multichannel end-to-end speech recognition},'' in \emph{ICASSP}, 2018.

\bibitem{anastasakos1996compact}
T.~Anastasakos \emph{et~al.}, ``{A compact model for speaker-adaptive training},'' in \emph{ICSLP}, 1996.

\bibitem{ochiai2014speaker}
T.~Ochiai \emph{et~al.}, ``{Speaker adaptive training using deep neural networks},'' in \emph{ICASSP}, 2014.

\bibitem{liao2013speaker}
H.~Liao, ``{Speaker adaptation of context dependent deep neural networks},'' in \emph{ICASSP}, 2013.

\bibitem{kitza2018comparison}
M.~Kitza \emph{et~al.}, ``{Comparison of BLSTM-Layer-Specific Affine Transformations for Speaker Adaptation},'' in \emph{INTERSPEECH}, 2018.

\bibitem{yu2013kl}
D.~Yu \emph{et~al.}, ``{KL-divergence regularized deep neural network adaptation for improved large vocabulary speech recognition},'' in \emph{ICASSP}, 2013.

\bibitem{li2018speaker}
K.~Li \emph{et~al.}, ``{Speaker adaptation for end-to-end CTC models},'' in \emph{SLT}, 2018.

\bibitem{meng2019speaker}
Z.~Meng \emph{et~al.}, ``{Speaker adaptation for attention-based end-to-end speech recognition},'' in \emph{INTERSPEECH}, 2019.

\bibitem{huang2020acoustic}
Y.~Huang \emph{et~al.}, ``{Acoustic model adaptation for presentation transcription and intelligent meeting assistant systems},'' in \emph{ICASSP}, 2020.

\bibitem{huang2015maximum}
Z.~Huang \emph{et~al.}, ``{Maximum a posteriori adaptation of network parameters in deep models},'' in \emph{INTERSPEECH}, 2015.

\bibitem{huang2016bayesian}
Z.~Huang \emph{et~al.}, ``{Bayesian unsupervised batch and online speaker adaptation of activation function parameters in deep models for automatic speech recognition},'' \emph{IEEE T AUDIO SPEECH}, 2016.

\bibitem{huang2020rapid}
Y.~Huang \emph{et~al.}, ``{Rapid RNN-T Adaptation Using Personalized Speech Synthesis and Neural Language Generator},'' in \emph{INTERSPEECH}, 2020.

\bibitem{sim2019personalization}
K.~C. Sim \emph{et~al.}, ``{Personalization of end-to-end speech recognition on mobile devices for named entities},'' in \emph{ASRU}, 2019.

\bibitem{meng2018speaker}
Z.~Meng \emph{et~al.}, ``{Speaker-invariant training via adversarial learning},'' in \emph{ICASSP}, 2018.

\bibitem{tsuchiya2018speaker}
T.~Tsuchiya \emph{et~al.}, ``{Speaker invariant feature extraction for zero-resource languages with adversarial learning},'' in \emph{ICASSP}, 2018.

\bibitem{price2014speaker}
R.~Price \emph{et~al.}, ``{Speaker adaptation of deep neural networks using a hierarchy of output layers},'' in \emph{SLT}, 2014.

\bibitem{swietojanski2015structured}
P.~Swietojanski \emph{et~al.}, ``{Structured output layer with auxiliary targets for context-dependent acoustic modelling},'' in \emph{INTERSPEECH}, 2015.

\bibitem{huang2015rapid}
Z.~Huang \emph{et~al.}, ``{Rapid adaptation for deep neural networks through multi-task learning},'' in \emph{INTERSPEECH}, 2015.

\bibitem{swietojanski2014learning}
P.~Swietojanski \emph{et~al.}, ``{Learning hidden unit contributions for unsupervised speaker adaptation of neural network acoustic models},'' in \emph{SLT}, 2014.

\bibitem{gillick1989some}
L.~Gillick \emph{et~al.}, ``{Some statistical issues in the comparison of speech recognition algorithms},'' in \emph{ICASSP}, 1989.

\bibitem{hu2022exploit}
S.~Hu \emph{et~al.}, ``{Exploiting Cross Domain Acoustic-to-articulatory Inverted Features for Disordered Speech Recognition},'' in \emph{ICASSP}, 2022.

\bibitem{povey2011kaldi}
D.~Povey \emph{et~al.}, ``{The Kaldi speech recognition toolkit},'' in \emph{ASRU}, 2011.

\bibitem{watanabe2018espnet}
S.~Watanabe \emph{et~al.}, ``{ESPnet: End-to-End Speech Processing Toolkit},'' in \emph{INTERPSEECH}, 2018.

\bibitem{wang23qa_interspeech}
H.~Wang \emph{et~al.}, ``{DuTa-VC: A Duration-aware Typical-to-atypical Voice Conversion Approach with Diffusion Probabilistic Model},'' in \emph{INTERSPEECH}, 2023.

\bibitem{luz2020alzheimer}
S.~Luz \emph{et~al.}, ``{Alzheimer’s Dementia Recognition through Spontaneous Speech: The ADReSS Challenge},'' in \emph{INTERSPEECH}, 2020.

\end{thebibliography}

\end{document}